\documentclass[final,5p,times,twocolumn]{elsarticle}
\usepackage[colorlinks]{hyperref}
\usepackage{url}
\usepackage{graphicx}
\usepackage{soul}
\usepackage{amssymb}
\usepackage{amsthm}
\usepackage[draft, inline, nomargin]{fixme}
\fxsetup{theme=color, mode=multiuser}
\FXRegisterAuthor{ab}{1}{\color{blue}AB}

\biboptions{sort&compress}

\journal{Nuclear Instruments and Methods in Physics}

\begin{document}

\begin{frontmatter}

\title{A prototype for SANDD: A highly-segmented pulse-shape-sensitive plastic scintillator detector incorporating silicon photomultiplier arrays}


\author[LLNL]{Viacheslav~A.~Li}\ead{li68@llnl.gov}
\author[LLNL]{Timothy~M.~Classen}
\author[LLNL]{Steven~A.~Dazeley}\ead{dazeley2@llnl.gov}
\author[UH]{Mark~J.~Duvall}
\author[UM]{Igor~Jovanovic}
\author[LLNL]{Andrew~N.~Mabe}
\author[LLNL]{Edward~T.~E.~Reedy}
\author[LLNL,UM]{Felicia~Sutanto}

\address[LLNL]{Lawrence Livermore National Laboratory, Livermore, CA 94550}
\address[UH]{Department of Physics and Astronomy, University of Hawai`i at M\={a}noa, Honolulu, HI 96822}
\address[UM]{Department of Nuclear Engineering and Radiological Sciences, University of Michigan, Ann Arbor, MI 48109}

\begin{abstract}
We report the first clear observation of neutron/gamma-ray pulse-shape sensitivity of a fully-instrumented 8 $\times$ 8 array of plastic scintillator segments coupled to two 5~cm $\times$ 5~cm 64-channel SiPM arrays as part of a study of the key metrics of a prototype antineutrino detector module designed for directional sensitivity. SANDD (a Segmented AntiNeutrino Directional Detector) will eventually comprise a central module of 64 elongated segments of $^{6}$Li-doped pulse-shape-sensitive scintillator rods, each with a square cross section of 5.4 mm $\times$ 5.4 mm, surrounded by larger cross section bars of the same material. The most important metrics with the potential to impact the performance of the central module of SANDD are 
neutron and gamma-ray pulse-shape sensitivity using silicon photomultipliers (SiPMs), particle identification via scintillator rod multiplicity, and energy and position resolution. As a first step, we constructed a prototype detector to investigate the performance of a central SANDD-like module using two 64-channel SiPM arrays and rods of undoped pulse-shape-sensitive plastic scintillator.

\end{abstract}

\begin{keyword}
pulse-shape discrimination \sep neutron imaging \sep reactor antineutrinos \sep  segmented plastic  scintillator \sep  non-proliferation \sep SiPM arrays

\end{keyword}

\end{frontmatter}

%\linenumbers

\section{Introduction}
In recent years, reactor antineutrino detection has become relatively straightforward, so much so that it is now possible to make multiple identical detectors with efficiencies that differ by only 0.2\%~\cite{DayaBayDetectorComp}. However, two outstanding problems relating to backgrounds remain. 
The first and most prevalent background is caused by fast cosmic-ray-induced neutrons, which can produce antineutrino-like correlated signals from proton recoils followed by neutron capture. Such correlated signals cannot be distinguished from true antineutrinos without particle identification. The second background concerns antineutrinos from neighboring reactors. Since most of today's detectors are at best minimally sensitive to antineutrino direction, it is impossible to determine if an antineutrino came from a reactor of interest or from other reactors that produce appreciable antineutrino flux at the detector location. In this paper, we describe the design and the results of a characterization study of a prototype module designed primarily for sensitivity to antineutrino direction and for reducing surface-related backgrounds. 

Reactor antineutrinos can be detected via inverse beta decay,
\begin{equation}
\bar\nu_{e} + p = e^{+} + n.
\end{equation}
The relative positions of the positron annihilation and neutron capture can reveal the direction of the incoming antineutrino~\cite{Vogel:1999zy}. 
The correlation is weak, however, producing a shift of only $\sim1.5$~cm on average. The neutron capture uncertainty, however, is somewhat larger, depending on the $^{6}$Li concentration. Large numbers of events and good position resolution for both interactions would be required~\cite{Apollonio:1999jg,Boehm:2000vp} to achieve sensitivity. The detection medium must contain hydrogen in order to facilitate this inverse beta decay reaction. For reasons associated with light output and attenuation length, most successful experiments are based on liquid organic scintillator~\cite{Boehm:2000vp,Reines1954,Cowan1956,Boehm:1980pm,Zacek:1986cu,Declais:1994su,Greenwood:1996pb,Apollonio:2002gd,Araki:2004mb,Bowden:2006hu,An:2012eh,Abe:2012tg,RENO:2015ksa,Boireau:2015dda,Ko:2016owz,Almazan:2018wln,Ashenfelter:2018zdm}. 
% LS detectors:
% Savannah River 1956 first observation~\cite{Reines1954,Cowan1956}
% KamLAND~\cite{Araki:2004mb}
% Savannah River 1996~\cite{Greenwood:1996pb}
% RENO~\cite{RENO:2015ksa}
% Palo Verde~\cite{Boehm:2000vp}
% CHOOZ~\cite{Apollonio:2002gd}
% Double CHOOZ~\cite{Abe:2012tg}
% DayaBay~\cite{An:2012eh}
% Nucifer~\cite{Boireau:2015dda}
% Neutrino-4~\cite{Serebrov:2017wml}
% SONGS~\cite{Bowden:2006hu}
% NEOS~\cite{Ko:2016owz}
% STEREO~\cite{Almazan:2018wln}
% PROSPECT~\cite{Ashenfelter:2018zdm}
% ILL~\cite{Boehm:1980pm}
% Bugey~\cite{Declais:1994su}
% Gosgen~\cite{Zacek:1986cu}
Most include a neutron capturing dopant such as Gd or $^{6}$Li to maximize the neutron-capture efficiency and reduce the neutron diffusion time to capture. Gd capture produces a gamma-ray shower, which results in a poor determination of the position of the capture. $^{6}$Li capture, however, produces a triton and an alpha, which in principle can allow for better position resolution. 

As discussed by the NuLat collaboration~\cite{Lane:2015alq}, there are advantages to using plastic rather than liquid scintillator. First is deployability; liquid scintillator must be blanketed with inert gas, can be chemically aggressive and flammable, requires careful engineering to adjust for atmospheric pressure changes, and can leak. Second, if antineutrino directional sensitivity is desired, plastic enables fine segmentation, which can contribute heavily to the key requirement of better position resolution.
Until now however, $^{6}$Li-doped plastic scintillators have not rivaled liquid scintillator in terms of pulse-shape discrimination (PSD) or light output performance. 
Most of the plastic detectors developed recently have employed non-homogeneous configurations with a separate neutron capturing material~\cite{Oguri:2014gta,Alekseev:2016llm,Abreu:2018pxg,Carroll:2018kad}. The exception is miniTimeCube~\cite{Li:2016yey}, which employed plastic scintillator doped with $^{10}$B; however, it did not have pulse-shape sensitivity and was a monolithic detector\footnote{miniTimeCube could potentially be a successful antineutrino detector design, but would require  a significant breakthrough in fast-timing high-channel-density readout electronics.}.

% PS detectors:
% Savannah River 1976 (with Cd absorber)
% Gd reflectors/wraps
% DANSS~\cite{Alekseev:2016llm}
% PANDA~\cite{Panda2017}, ~\cite{Oguri:2014gta}
% SOLID~\cite{Abreu:2018pxg}, old Abreu:2017bpe
% Vidaar~\cite{Vidaar2018} or ~\cite{Carroll:2018kad}
% Except :
% miniTimeCube
% NuLat

An above ground antineutrino detector must differentiate rare correlated positron and neutron-capture events from more common uncorrelated gamma rays and neutron-induced proton recoils. If this can be done with minimal shielding from cosmic rays, detector technology may be able to transition toward smaller and more mobile concepts of interest to safeguards organizations. The segmented antineutrino directional detector (SANDD) will be constructed of a new form of $^{6}$Li-doped pulse-shape-sensitive plastic scintillator. 
The choice of $^{6}$Li as a dopant is motivated by its high capture cross-section for thermal neutrons, the localized energy deposition of its capture products, and its production of distinguishable differences in pulse shape.
In organic scintillators, neutron interactions such as proton recoils and neutron capture on $^{6}$Li produce higher concentrations of molecular triplet states, which yield a longer light pulse than electrons.
%~\cite{ZAITSEVA201288,ZAITSEVA2013747}.
Pulse-shape sensitivity provides a crucial ability to distinguish neutron elastic-scattering and capture interactions from those of gamma-ray scattering. 
Small samples of this scintillator have been manufactured in recent years~\cite{ZAITSEVA201288,ZAITSEVA2013747,MABE201680}. However, for construction of SANDD, these materials need to be manufactured at larger scale.

\begin{figure}[ht]
\centering\includegraphics[width=0.65\linewidth]{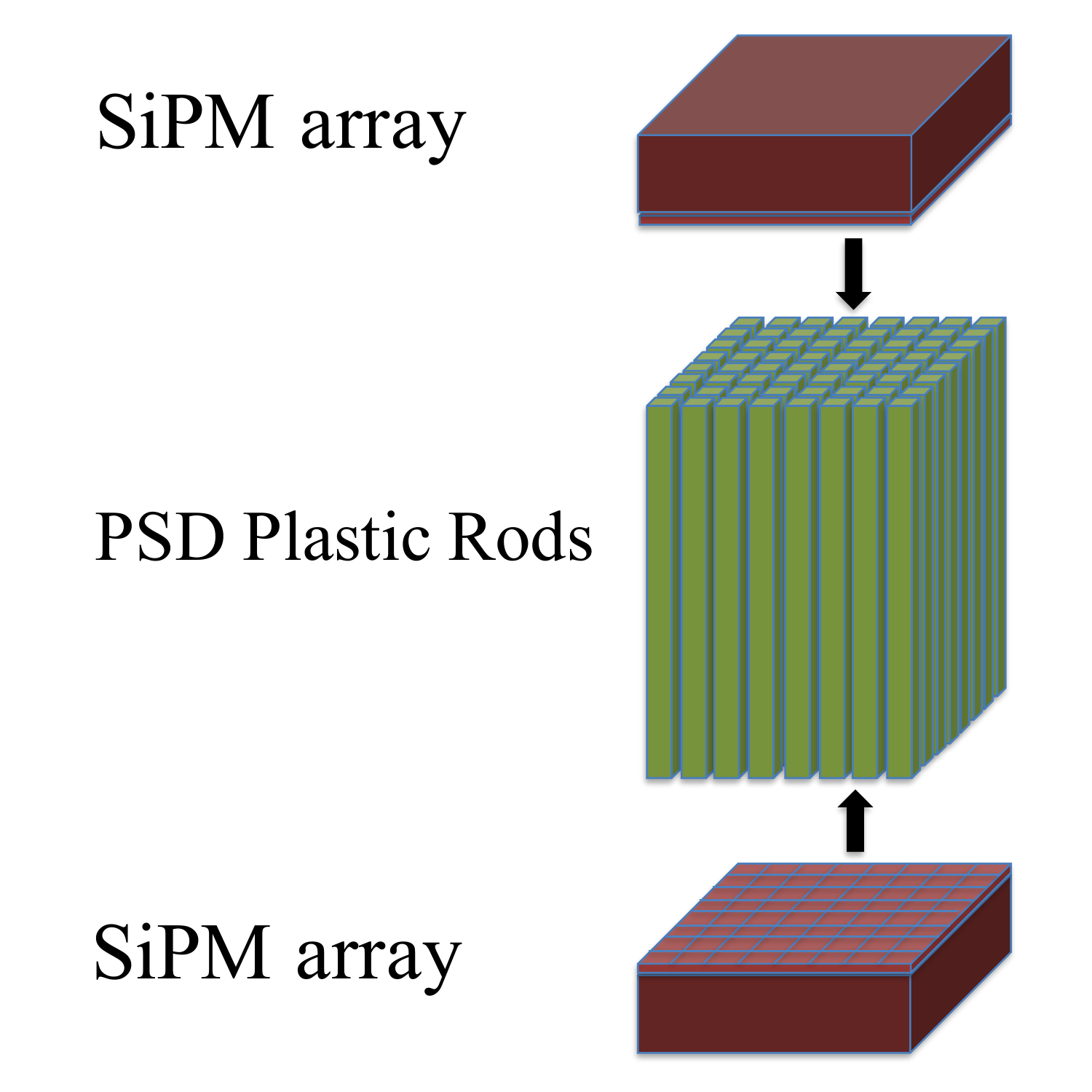}
\caption{An illustration of the prototype module characterized in this work. The module has similar characteristics to the SANDD central module design, featuring an $8\times 8$ array of undoped PSD plastic-scintillator rods. Signals from the scintillator segments are read individually at each end by the SiPM arrays.} 
\label{fig:detector_diagram}
\end{figure}

The prototype detector described here was built with an early form of PSD plastic scintillator without the incorporation of $^{6}$Li. The ability to manufacture undoped PSD plastic preceded the $^{6}$Li-doped plastic by about 6 months.  
The present prototype was built in order to test the data acquisition system (DAQ) and to study some of the key metrics that will impact the performance of the final SANDD module. These include pulse-shape sensitivity when using a dual 64-channel silicon photomultiplier (SiPM) readout, light transport along the scintillator rods, particle identification from segment multiplicity, and energy and position resolution. The detector constructed for this test incorporates 64 rods of PSD plastic scintillator of size $5.4\;\mathrm{mm}\times5.4\;\mathrm{mm}\times112\;\mathrm{mm}$ arranged in an $8\times8$ array. 
One 64-pixel SiPM array (SensL J-60035 series, 50.44~mm~$\times$~50.44~mm) is mounted at each end of the scintillator-rod array. 
The J-series SiPMs used in this work have 22,292 35-$\mu$m microcells per pixel and typically are operated at a bias of between 2 and 6 V above the breakdown voltage of $24.5 V$. The peak efficiency at 420~nm is between $38 - 50\%$ and dark rates of between $50-150$~kHz/mm$^2$, depending on the bias. 
Each pixel had an active area of 6.13~mm~$\times$~6.13~mm, with a 0.2-mm gap between each pixel.

More details on this emerging technology can be found in the original reports~\cite{Roncali2011,SensL} and references therein.
A schematic that illustrates the general design features of the prototype is shown in Fig.~\ref{fig:detector_diagram}. 
%According to the specifications of the SiPM arrays provided by SensL, for the 2.5--6~V range of overvoltage:  the 420-nm photon detection efficiency is in the range of 38--50\%, gain --- 2.9$\times$10$^6$--6.3$\times$10$^6$, dark count rate --- 50--150~kHz/mm$^2$. The spectral range is in 200--900~nm, with a peak at 420~nm. The breakdown voltage is in the range of 24.2--24.7~V, and the maximum overvoltage is 6~V, which correspond to the maximum operating (or bias) voltage of 30.7~V. 
%Temperature dependence of breakdown voltage --- 21.5~mV/K.
%Each pixel has 22,292 35-$\mu$m microcells on a 6.13~mm~$\times$~6.13~mm active area ($\sim$73\% fill factor), with a 0.2-mm gap between each pixel. 
%More details on this emerging SiPM technology can be found in the original reports~\cite{Roncali2011,SensL} and references therein.

For antineutrino detection, the segmentation afforded by the use of square cross-section rods of scintillator directs photons up and down towards the two SiPM arrays, allowing for position reconstruction in the plane perpendicular to the direction of the scintillator rods. 
In the following, the $z$ axis of the detector is defined as parallel to the scintillator rods, while the $x\textrm{-}y$ plane is defined as perpendicular to the scintillator rods. 
The other advantage of segmentation is the ability to determine particle type based on the number of segments that have energy deposited in them. For neutron-induced proton recoil or neutron capture on $^{6}$Li, the resulting protons, tritons, and alphas are predicted to deposit the bulk of their energy within a single rod. 
Compton-scattered electrons or inverse-beta-decay positrons, however, are predicted to deposit their energy over a longer range, often producing light in two or more contiguous rods. The annihilation gamma rays from a positron may travel a significant distance before depositing energy in the detector. 
The exploitation of differences in range and rod multiplicity for neutron interactions compared to other particles (gammas, betas, muons) adds another independent discriminant to differentiate rare antineutrino interactions from backgrounds. 
This complements pulse-shape discrimination and positron-neutron position and time correlation. We report on measurements of particle identification via multiplicity in Section~\ref{rodMultiplicity}.

In this paper, we focus on the following performance metrics:
\begin{enumerate}
	\item Neutron/gamma pulse-shape discrimination capability
    \item Light transport and position resolution along the scintillator rods
    \item Relative energy resolution for different configurations
    \item Particle identification via rod multiplicity 
\end{enumerate}
Significantly, for SANDD and other similar segmented-detector designs, we report possibly the first observation of PSD in a segmented plastic scintillator array coupled to 64-channel SiPM arrays.

\section{The detector}

The plastic scintillator used in the prototype was fabricated with the aim of developing meter-scale PSD plastic scintillator. A full description of the development of large-scale PSD plastic scintillators suitable for antineutrino-detection applications will be described in an upcoming publication. The scintillator fabrication process is summarized as follows. Styrene and divinylbenzene (DVB) were distilled and degassed immediately before use. In a nitrogen-purged glovebox, appropriate quantities of {\it 2,5-diphenyloxazole} (PPO), {\it 1,4-bis(2-methylstyryl)benzene} (bis-MSB), and DVB were added to styrene to make mass fractions of 30\%, 0.2\%, and 5\%, respectively. An amount of 0.05\% {\it 1,1-bis(tert-butylperoxy)-3,3,5-trimethylcyclohexane} (L-231 initiator) was added, and then the resulting solution was poured into an aluminum mold. The mold was sealed, removed from the glovebox, and placed in a nitrogen-purged oven for polymerization. The polymerization temperature profile was optimized to eliminate air bubbles and maximize polymerization completion. After polymerization, the solid scintillator block was removed from the mold and machined into sixty-four rods of size $5.4\;\mathrm{mm}\times5.4\;\mathrm{mm}\times112\;\mathrm{mm}$. The resulting rods were polished by standard optical polishing techniques.
%PPO: 2,5-diphenyloxazole
%Bis-MSB: 1,4-bis(2-methylstyryl)benzene

\begin{figure}[ht]
\centering\includegraphics[width=1.0\linewidth]{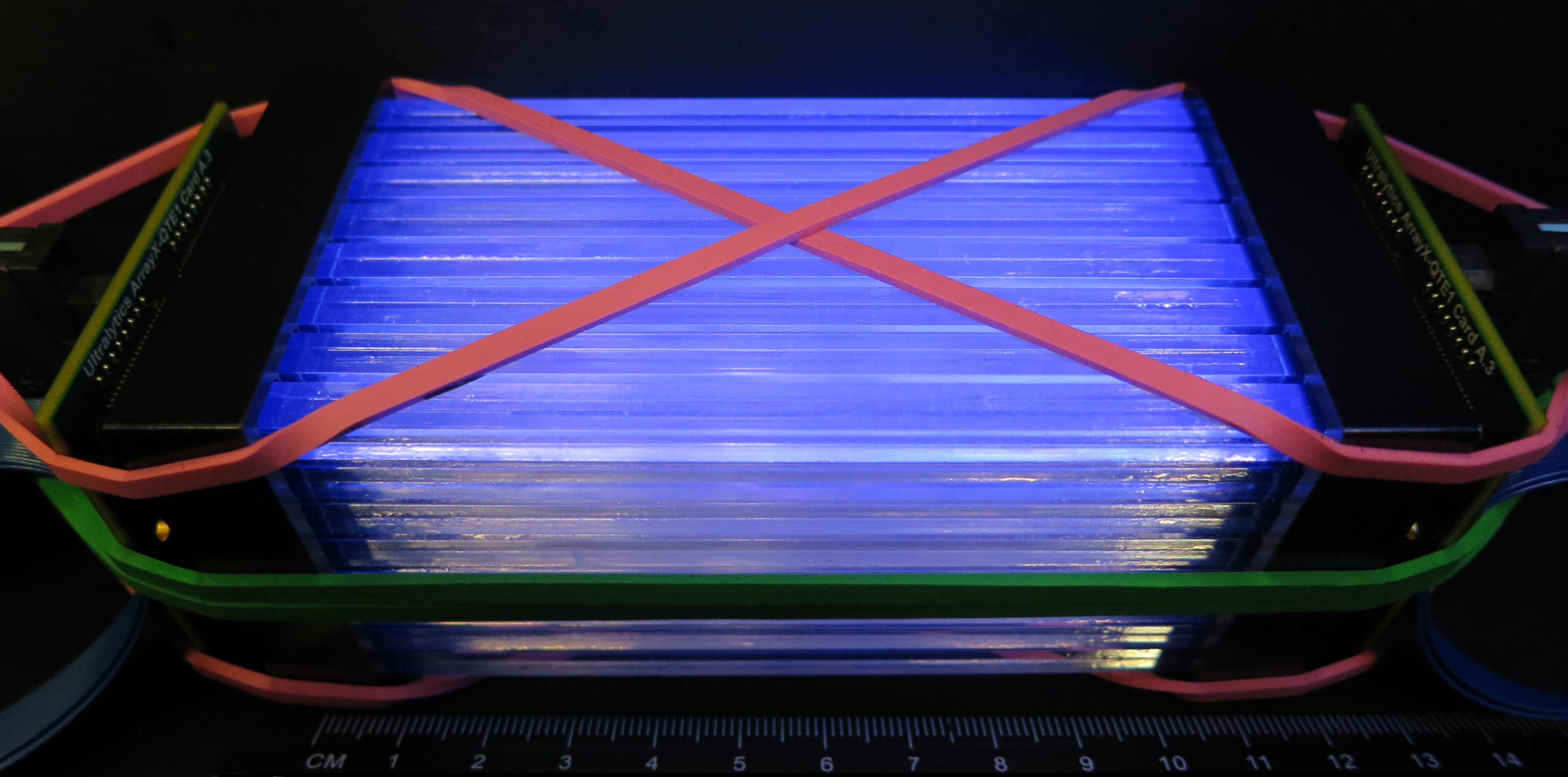}
\caption{The prototype detector's full assembly: 64 PSD plastic-scintillator rods are ``sandwiched'' in between two 64-channel SiPM arrays.}
\label{fig:full_assembly}
\end{figure}

%The scintillator rods in two rod-alignment frames  and SiPM arrays are shown in Fig.~\ref{fig:full_assembly} and were placed in a light-tight enclosure. 
The scintillator rods were supported by an 8 x 8 square plastic frame that was fitted to each SiPM. The frame maintained the alignment of each scintillator rod with its own SiPM channel while keeping them all separated. Two rod-alignment frames, the scintillator rods and the SiPM arrays are shown together in Fig.~\ref{fig:full_assembly} and were placed in a light-tight enclosure.

\begin{figure}[ht]
\centering\includegraphics[width=1.0\linewidth]{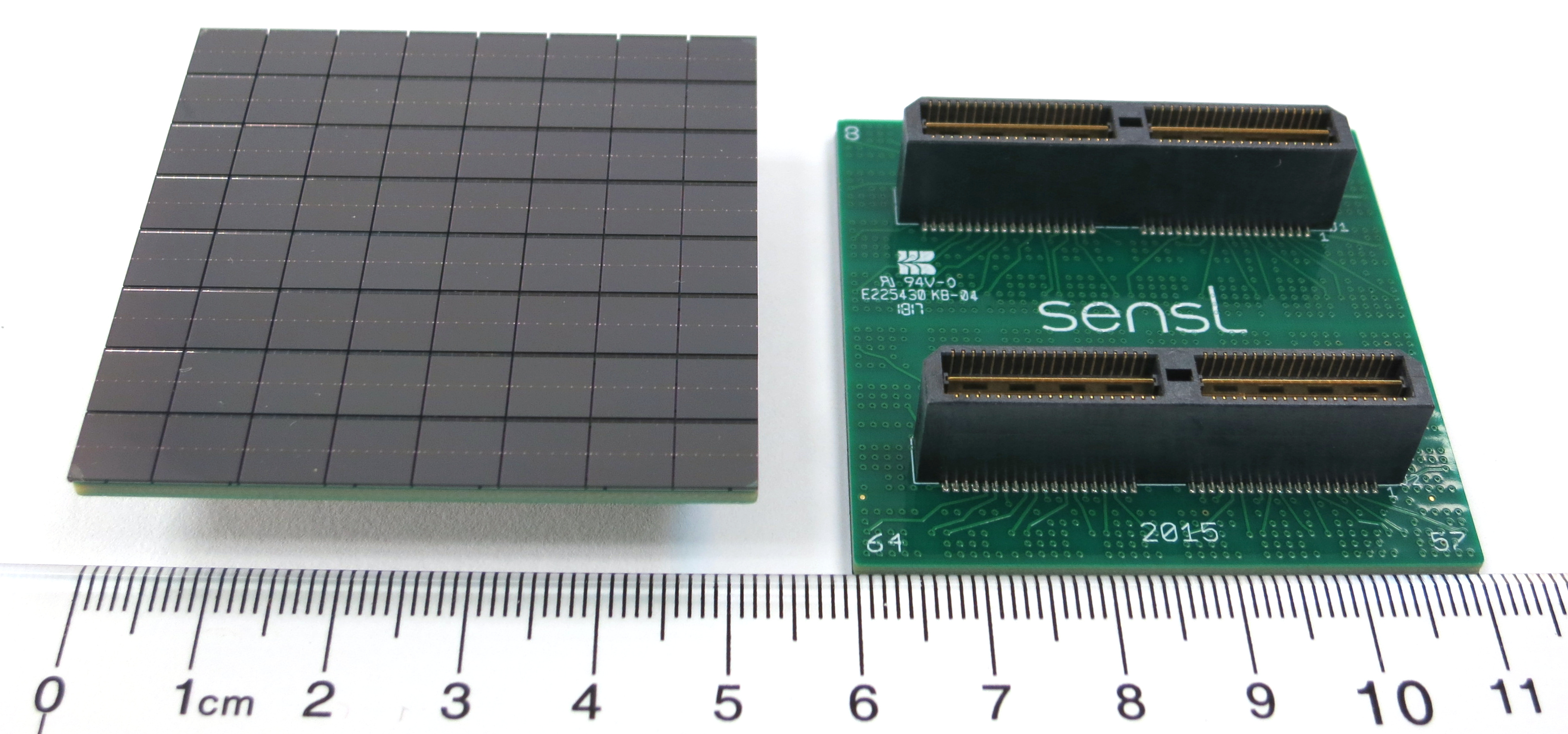}
\caption{Two SiPM $8 \times 8$  arrays (front and back) used in the experiment.}
\label{fig_SiPM_array}
\end{figure}

The plastic scintillator rods were optically coupled to the SiPM arrays using BC-630 silicone optical grease.
The two SiPM arrays, shown in Fig.~\ref{fig_SiPM_array}, were individually powered. The SiPM operating voltage  was set at 30~V for most of the non-amplified tests and at 27~V for tests with the 10$\times$-gain amplifiers. The combination of the amplification and lower bias voltage was chosen to reduce noise in the SiPM array signals.

\begin{figure}[ht]
\centering\includegraphics[width=1.0\linewidth]{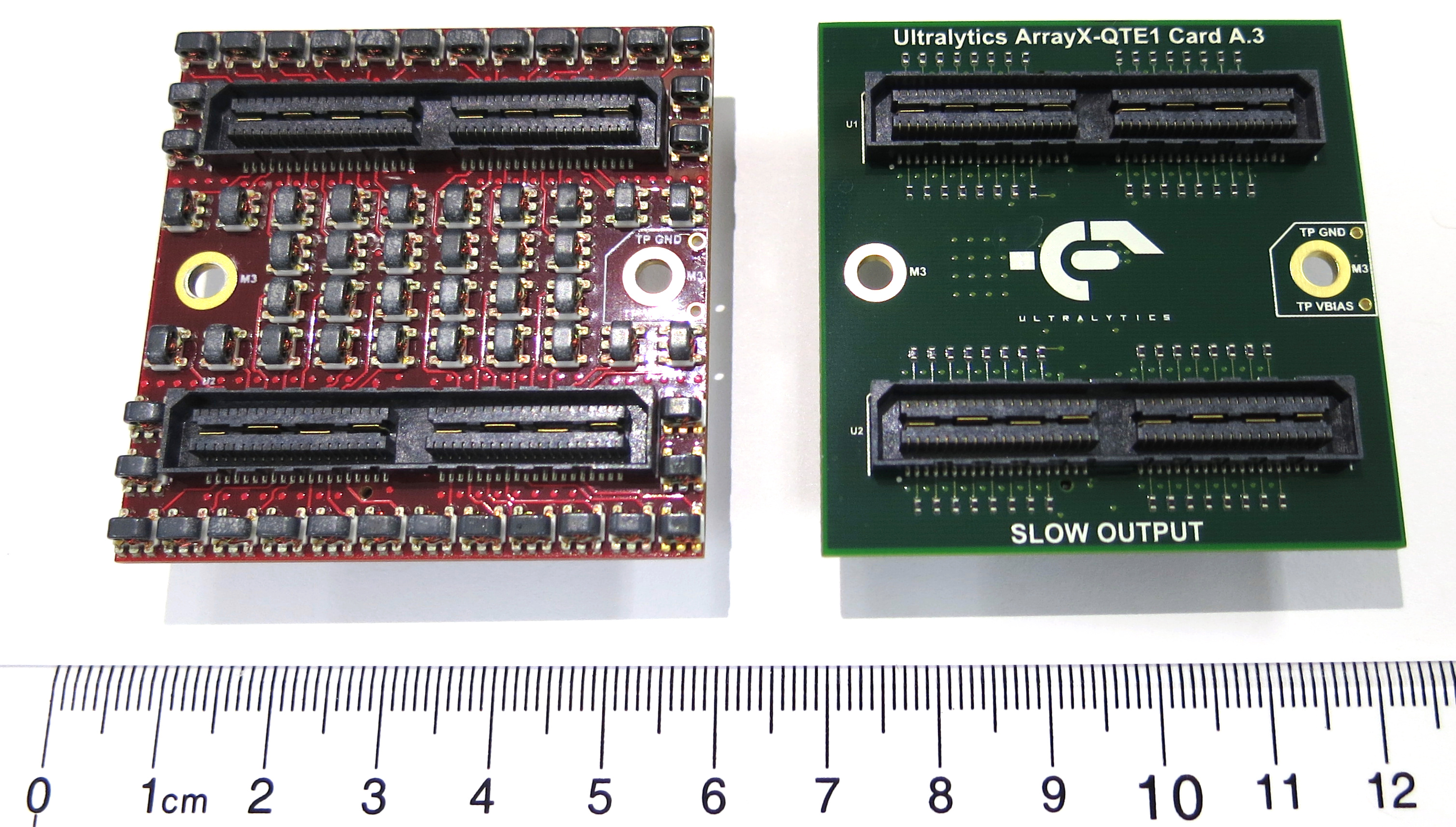}
\caption{Two types of interconnect SiPM cards: with fast-output (left) and slow-output (right). The SensL array-J SiPMs are connected to the two connectors on the card. The fast output card has 64 balun transformers (visible in the left module).}
\label{fig:Ultralytics_cards}
\end{figure}

One of the key components of the detector is an interconnect board for reading out the individual signals from all 64 channels of each SiPM array, while preserving the pulse-shape information from the PSD plastic photon-emission profile. 
The first such boards, obtained from Ultralytics, were designed to generate the fast differential output of the SiPM array. 
Although these boards have good energy resolution and fast timing, the differential output does not preserve the pulse-shape differences present in the  scintillator luminescence. 
A second iteration of this board, designed for this experiment and also from Ultralytics, produced a ``slow-output'' or non-differential readout.
The two boards are shown in Fig.~\ref{fig:Ultralytics_cards}. The fast-output card has balun %(balanced-unbalanced) 
transformers (64 in total, one per channel), resulting in output pulses with short ($\sim$10~ns) decay time, long overshoot below baseline, and little PSD information.
In contrast, the pulses from individual SiPM-array pixels coupled with the slow-output cards have a decay time of approximately 200~ns.

\begin{figure}[ht]
\centering\includegraphics[width=1.0\linewidth]{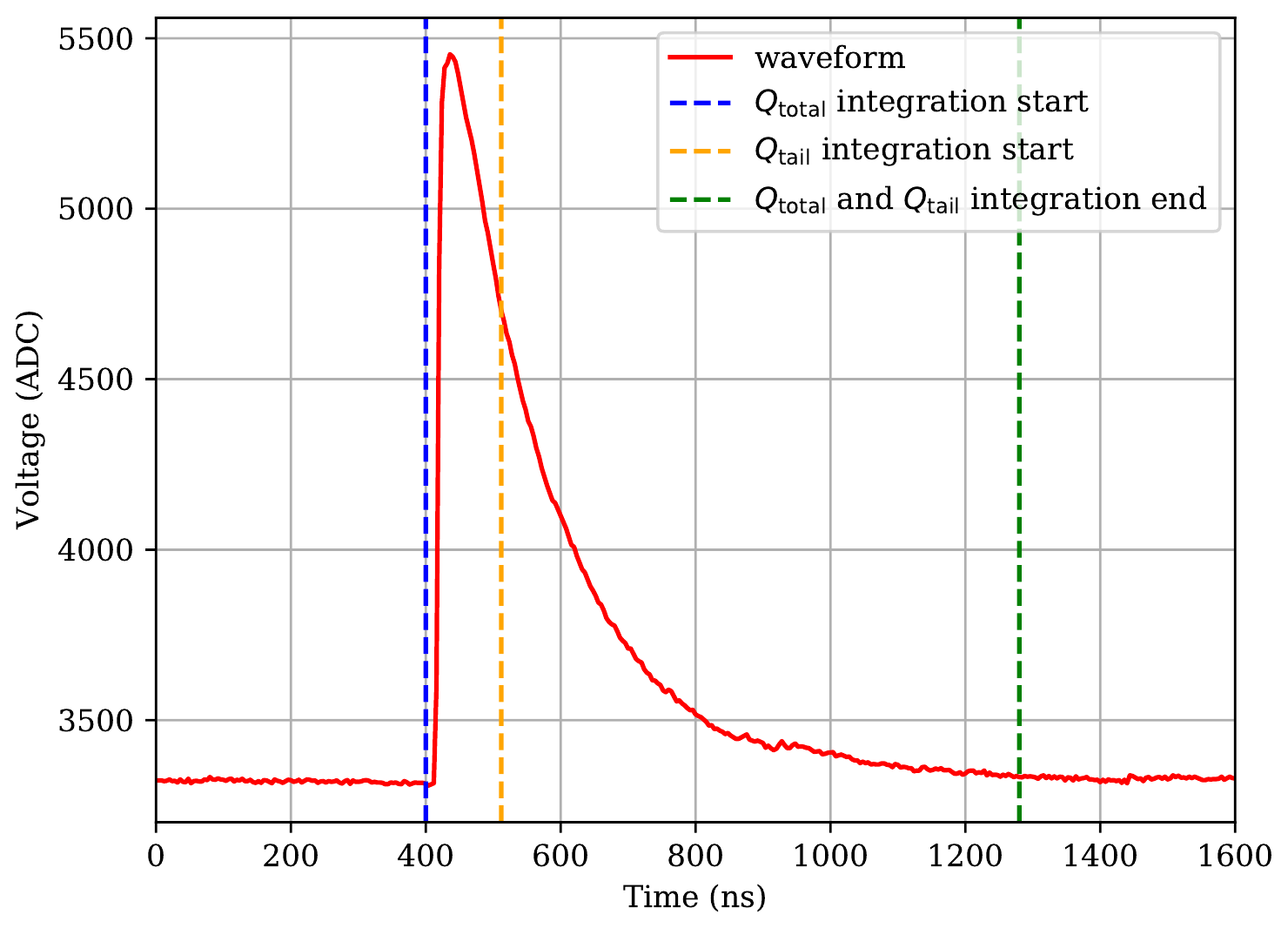}
\caption{A typical slow-output waveform from one of the SiPM-array pixels acquired using 250-MHz Struck SIS 3316 digitizer module, with 10$\times$ amplification, before pedestal subtraction. }
\label{fig:sample_waveform_slowoutput}
\end{figure}

The DAQ system comprises eight 16-channel VME Struck SIS3316 digitizer modules (250~MS/s, 14~bit, 5~V  dynamic range), providing 128 individually-triggered full-waveform channels in total.
A simple trigger implementation was chosen so that any one of the 128  channels over threshold triggers the recording of a 400-sample-long waveform from that channel. A threshold of approximately 0.1 MeVee was employed for all of the data presented in the following. This simple trigger allows for either multi- or single-rod events to be recorded for later analysis. Waveforms are sent to disk as each digitizer buffer fills. An event builder was implemented in software to time-sort and combine the waveforms that occur within 1~$\mu$s as single ``events''. An example of a 400-sample (1600 ns) waveform is shown in Fig.~\ref{fig:sample_waveform_slowoutput}. 
The integration times used to determine the waveform pulse shape are adjustable and were optimized using calibration data from neutron and gamma-ray sources (see Section~\ref{calibration}).
In the analysis, an energy deposit must trigger the SiPM pads on both ends of the scintillator rod for it to be considered a real event trigger.

The DAQ is controlled from a Linux desktop machine via a fiber-optic VME-to-PCI interface (Struck SIS 3100/1100). % with a Struck PCI fiber-optic card.
The raw data from the digitizers is stored in ROOT format~\cite{Brun:1997pa}, and then processed by a series of ROOT routines to handle the event building and analysis. 

\section{Calibration and analysis}
\label{calibration}

Each waveform saved to disk consists of 400 samples over 1600 ns. 
The first step in the analysis procedure is the calculation of the pedestal. The pedestal for each channel is simply an average over the first 60 samples (240 ns) of each waveform. A PSD-analysis routine is subsequently applied to determine if the pulse shape is neutron-like or gamma-ray-like. Since the light output from the scintillator due to neutron interactions produce longer light pulses, we employ a simple $Q_{tail}/Q_{total}$ metric for each event as follows:
\begin{equation}
    \frac{Q_{\mathrm{tail}} }{ Q_{\mathrm{total}} }
        =
    \frac{
    \sum \limits_{i}^{rods} \sqrt{ Q_{\mathrm{tail}}^{A\; i} Q_{\mathrm{tail}}^{B\; i} } 
    }{
    \sum \limits_{i}^{rods} \sqrt{ Q_{\mathrm{total}}^{A\; i} Q_{\mathrm{total}}^{B\; i}}
    }
\end{equation}
where the sums are over the rods with triggered pixels on both ends (both SiPM A and SiPM B sides) and exclude rods with either A or B triggered end; the tail and total charges ($Q_{tail}^i$ and $Q_{total}^i$) for each channel are integrated over the following time intervals  (see Fig.~\ref{fig:sample_waveform_slowoutput}):

\begin{equation}
\label{IntegrationTiming}
\ \ \ \ 0\ \mathrm{ns} \ \leq \ T_{\mathrm{total}} \ \leq \ 880 \ \mathrm{ns} 
\end{equation}
\begin{equation}
\label{IntegrationTiming2}
112\ \mathrm{ns} \ \leq \ T_{\mathrm{tail}} \ \ \;\! \leq \ 880 \ \mathrm{ns},
\end{equation}
where $T = 0$~ns is defined as the start time of the charge integration, 20 ns (5 samples) before the leading edge of each pulse. These integration times were determined from an optimization analysis described in Section~\ref{PSD}. 

\begin{figure}[ht]
\centering\includegraphics[width=1.0\linewidth]{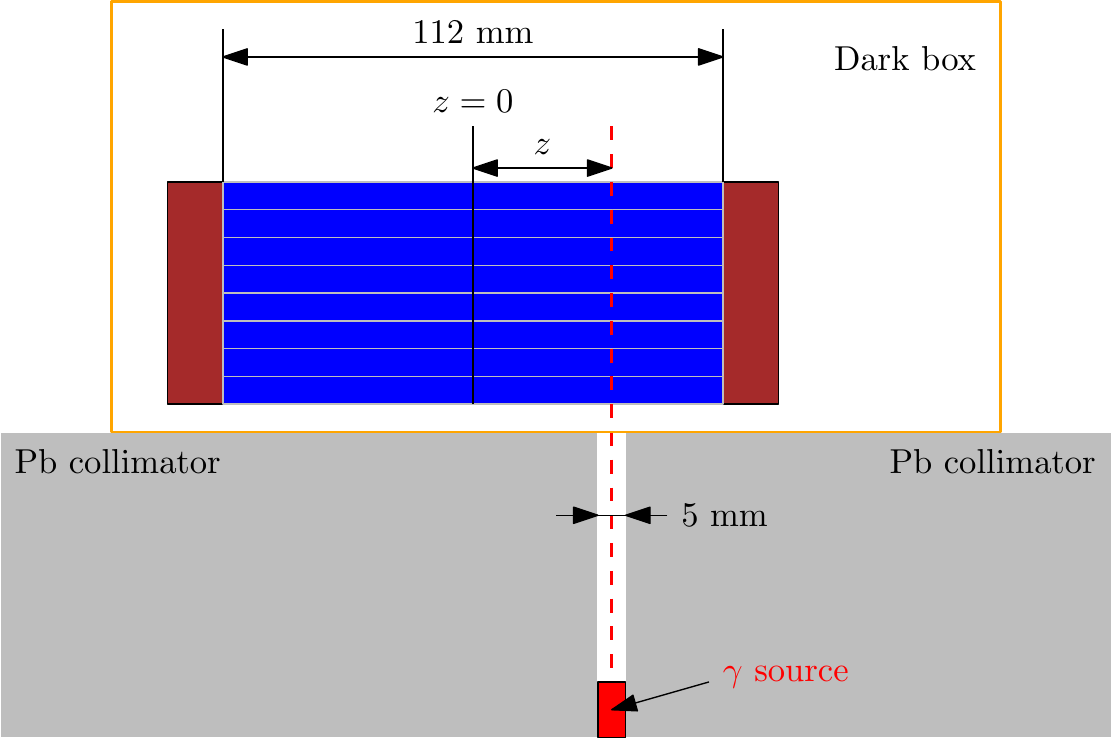}
\caption{A top-view diagram of the setup with collimated ${}^{137}$Cs source, used for energy/position calibration. The vertical slit is 5-mm wide. The source is 15~cm away from the enclosure and $\sim$18~cm away from the closest rod.}
\label{fig_Cs_collimated}
\end{figure}

To ensure uniformity of response from all 128 channels in contact with scintillator, each 64-channel SiPM array was evenly illuminated with pulses of light from an LED, and then the gain was corrected to equalize all the channels. The scintillator bars were subsequently coupled to the SiPMs and a collimated beam of $^{137}$Cs 662~keV gamma rays was directed at the scintillator volume at a position corresponding to $z=0$ (the center, see Fig.~\ref{fig_Cs_collimated}). The charge spectra from each of the channels were obtained and the Compton edges for each channel identified and assigned a corresponding charge value. The differences among the channels were attributed to differences in coupling efficiency between the scintillator rods and the SiPM pads. The apparent signal intensities of each channel were then corrected relative to the SiPM channel with the largest Compton edge charge value, so that all channels generated an equivalent response.
Following calibration, the measured charge at both ends of each scintillator rod can be correlated to the event energy by assuming that the signal measured depends exponentially on the distance along the rod as follows~\cite{Nelson:2011ux}:
\begin{equation}
\sqrt{E_A E_B} = 
    \sqrt{(E e^{-z / \alpha_{eff}})(E e^{z / \alpha_{eff}})} = E,
    \label{EqEnergy}
\end{equation}
where $E_A$ and $E_B$ are energy depositions (in arbitrary charge units) as measured by SiPM A and B, respectively. 
%So long as the light output detected at each end of each rod follows an  exponential dependence on the distance to the event, 
The  geometric mean of the light output recorded at the two ends of the rod is proportional to deposited energy and does not depend on the $z$---coordinate. 

\begin{figure*}[ht]
\centering\includegraphics[width=1.0\textwidth]{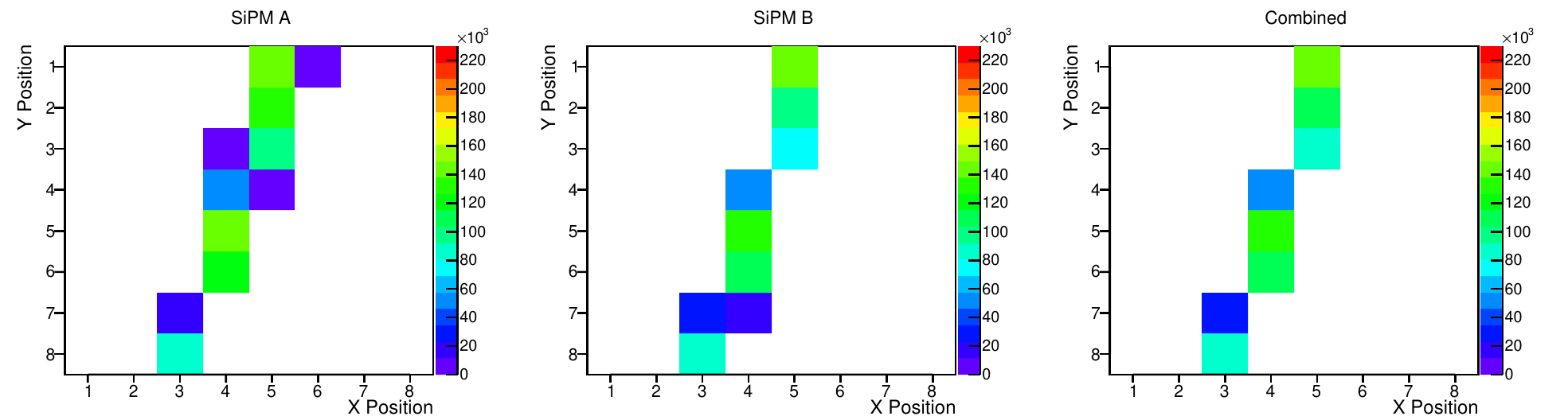}
\caption{An event attributed to a cosmogenic muon passing through the detector and mapped onto both SiPM A and B from the reference frame of the scintillator rods (see Fig.~\ref{fig_muon_CAD} for an example). The left and center panels show the charge collected on SiPM array A and B, respectively; the right panel shows the combined signal where the total charge in each scintillator rod is combined according to Eq.~\ref{EqEnergy}.}
\label{fig_muon}
\end{figure*}

\begin{figure}[ht]
\centering\includegraphics[width=.55\linewidth]{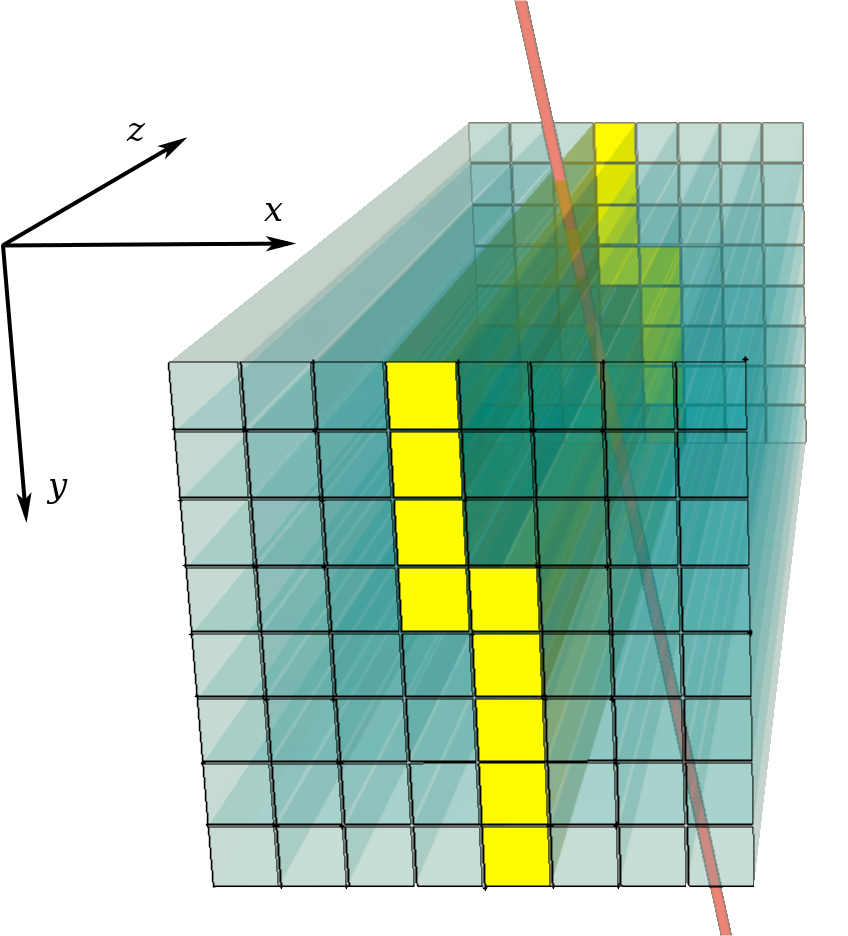}
\caption{A visualization of a cosmogenic muon (shown in red) passing through the detector. The rods with energy deposited by the muon are shown in yellow, while the rods missed by the muon are shown in green.}
\label{fig_muon_CAD}
\end{figure}

Following calibration, the detector was ready for physics events. For instance, Fig.~\ref{fig_muon} shows an example of the detector response to a cosmogenic-muon candidate, as mapped onto SiPM A and B from the point of view of a coordinate system defined by the arrangement of the scintillator rods. Fig.~\ref{fig_muon_CAD} shows the coordinate system used and a visualization of a hypothetical muon track through the scintillator rods. 

\begin{figure}[ht]
\centering\includegraphics[width=1\linewidth]{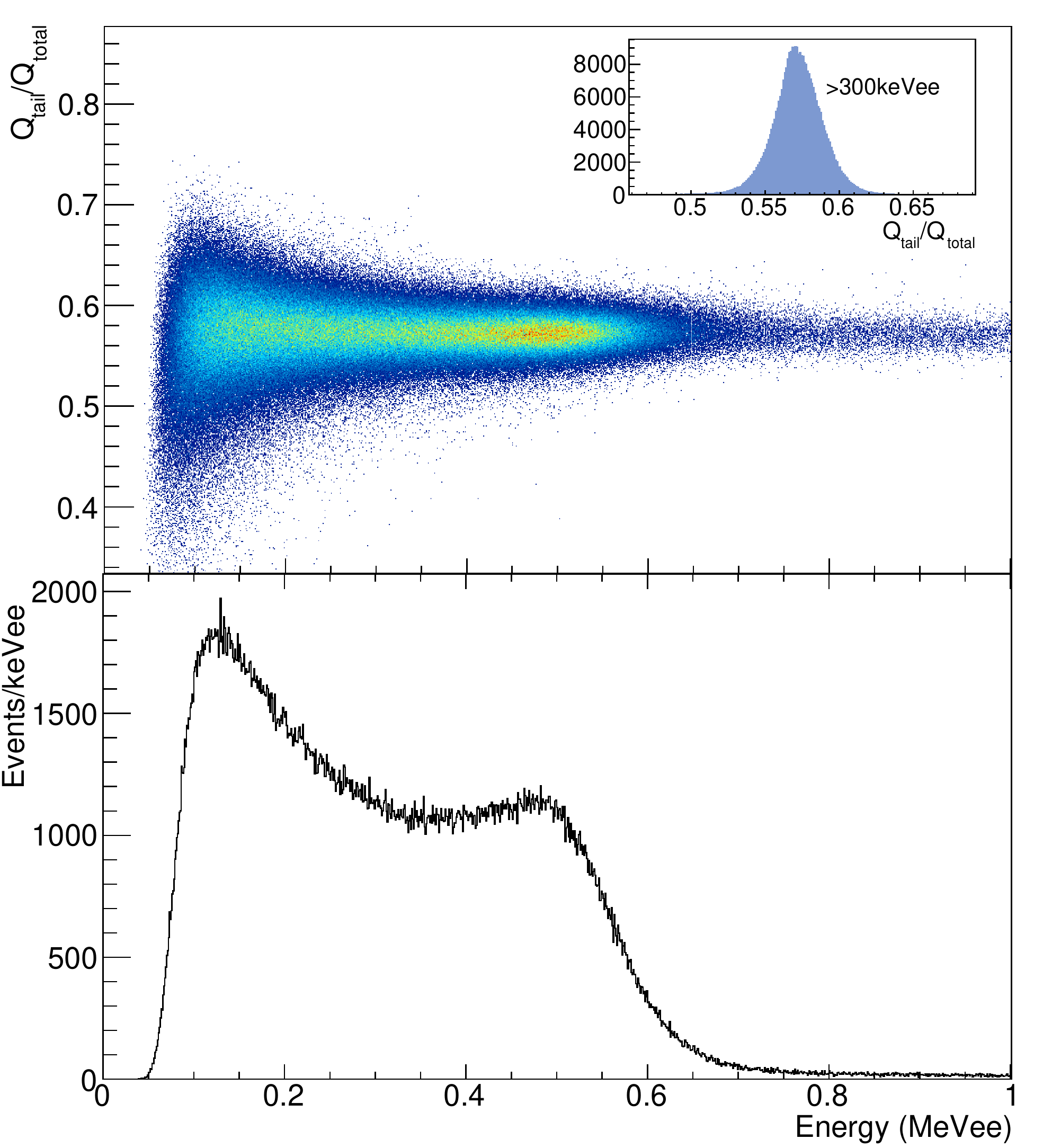} %pspPlotCs4_new.pdf
\caption{Top: a distribution of $Q_{\mathrm{tail}}/Q_{\mathrm{total}}$ vs Energy for $^{137}$Cs source. Bottom: spectrum (x-axis projection of top plot). Inset: a distribution of $Q_{\mathrm{tail}}/Q_{\mathrm{total}}$ with total energy above 300~keV$_{\mathrm{ee}}$.}
\label{pspCs}
\end{figure}

Fig.~\ref{pspCs} shows the PSD and the total energy detected in the 64-rod module for an uncollimated $^{137}$Cs source following calibration. 
The Compton edge is situated at approximately 500~keV$_{\mathrm{ee}}$. To define the conversion factor between charge and energy, we fit a Gaussian to the $^{137}$Cs spectrum at the Compton-edge region; we defined the mean as 477~keV$_{\mathrm{ee}}$. The ratio of the standard deviation to the mean is referred to as the energy resolution ($\sim 16\%$ at the $^{137}$Cs Compton edge). %, and depends on the configuration of the setup (e.g. use of Teflon wrapping, as discussed in Section~\ref{sub_teflon_wrap}). 
Also shown is the pulse shape quantified in terms of $Q_{tail}/Q_{total}$ as a function of energy, where energy is defined for any particle as the equivalent energy of an electron (MeV$_{\mathrm{ee}}$), determined using Equation~(\ref{EqEnergy}). The equivalent PSD plot for an uncollimated $^{252}$Cf source is presented in Section~\ref{PSD}.

\begin{figure}[ht]
\centering\includegraphics[width=1.0\linewidth]{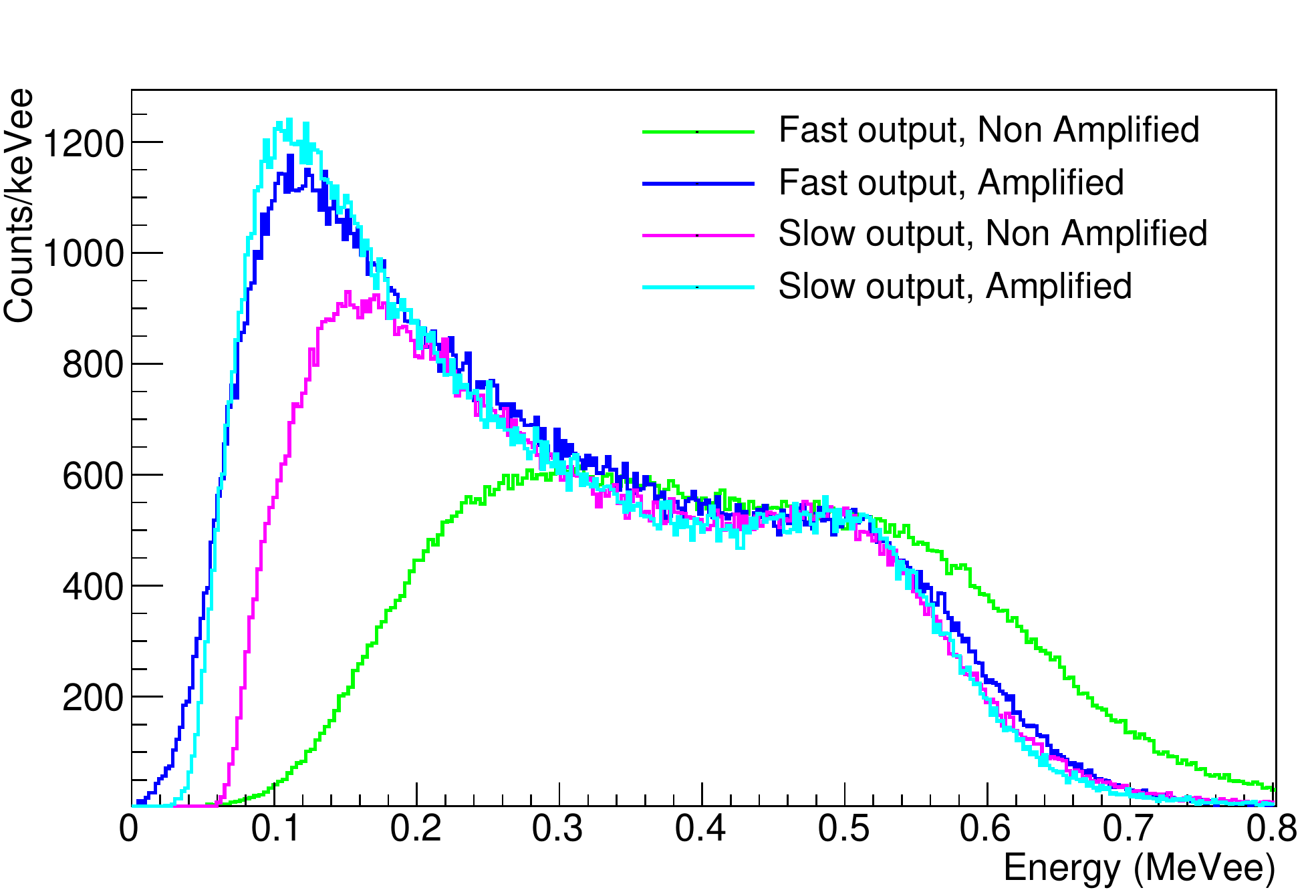}
\caption{Comparison of ${}^{137}$Cs spectra for different readout configurations. These tests were performed without teflon wrapping.}
\label{fig_Cs_spectra_comparison_1}
\end{figure}

\begin{figure}[ht]
\centering\includegraphics[width=1.0\linewidth]{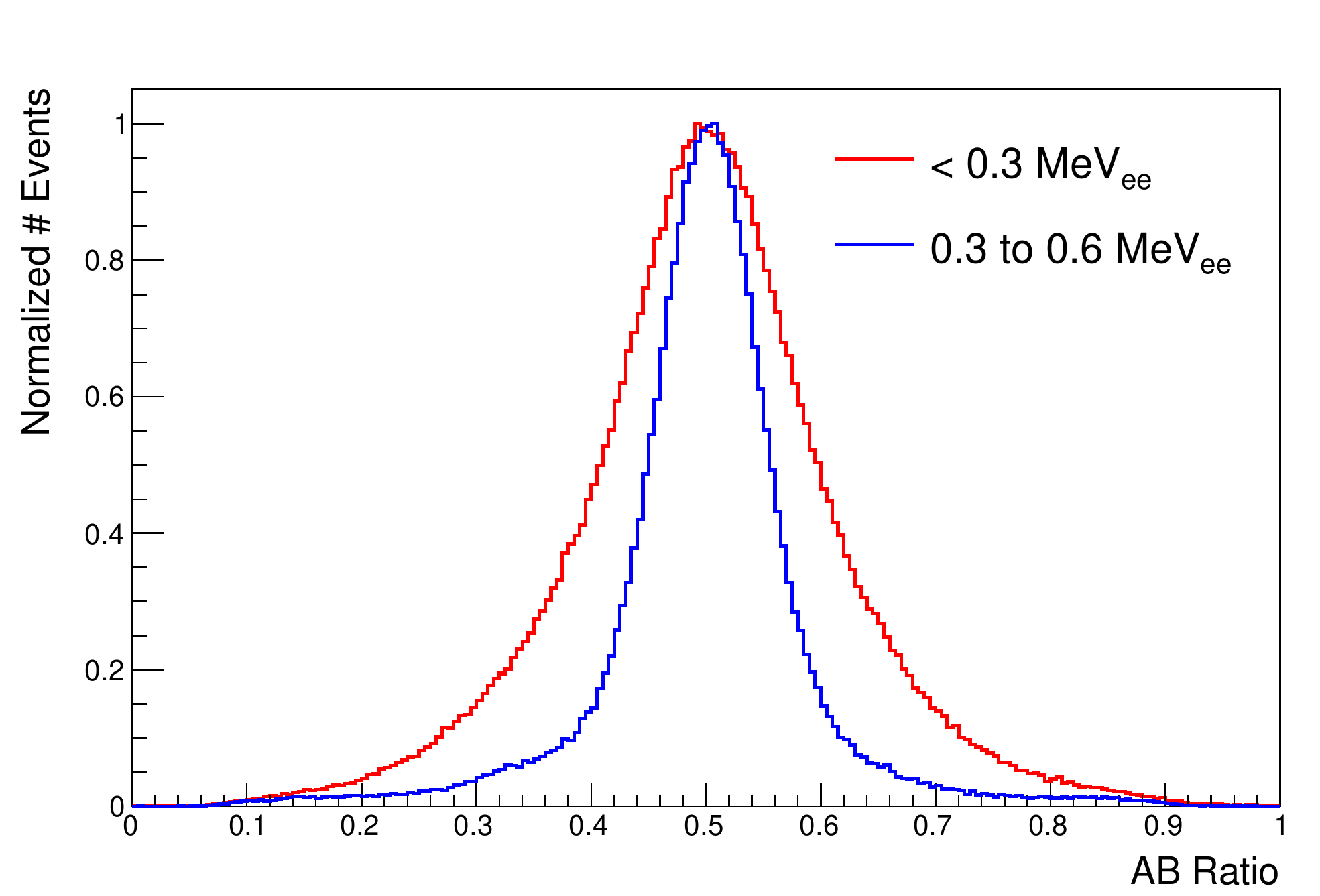}
\caption{A plot of the charge observed in SiPM A as a fraction of the summed charge observed from A and B. Two energy ranges are plotted to illustrate the improvement in AB ratio resolution that occurs with increasing energy: 0 to 0.3 MeV$_{\mathrm{ee}}$ and 0.3 to 0.6 MeV$_{\mathrm{ee}}$. Both curves were scaled to the same peak height  to emphasize the improvement in resolution.}
\label{fig_AB_ratio_vs_energy}
\end{figure}

\begin{figure}[ht]
\centering\includegraphics[width=1.0\linewidth]{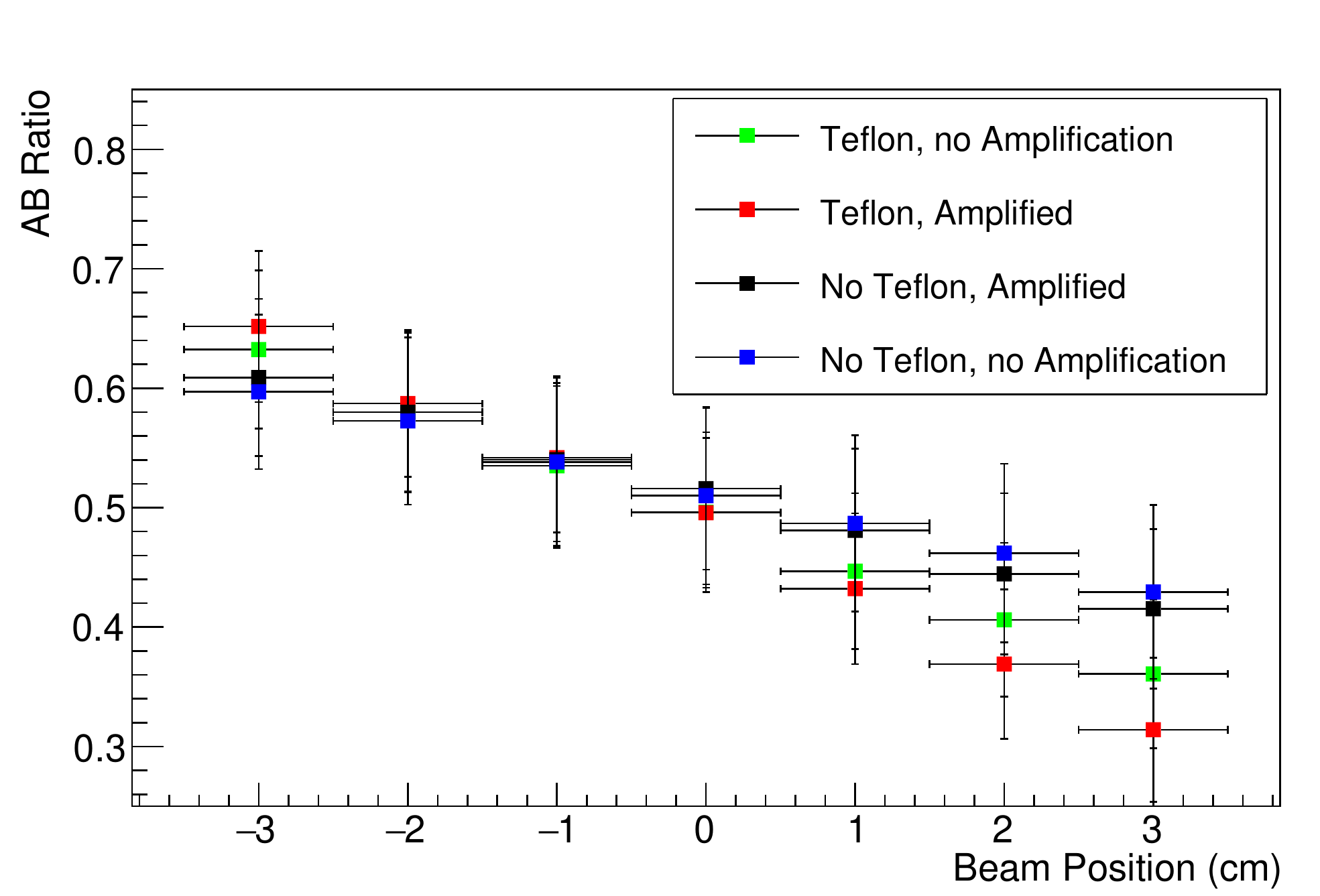}
\caption{The AB ratio as a function of collimated beam $z$ position (see Fig.~\ref{fig_Cs_collimated} and Eq.~\ref{eq_AB_ratio}) for four different detector configurations. The error bars shown here indicate the width of the $R_{AB}$ distribution (the event-by-event uncertainty), not the uncertainty in the mean at each beam position.}
\label{fig_AB_ratio_vs_position}
\end{figure}

\begin{figure}[ht]
\centering\includegraphics[width=1.0\linewidth]{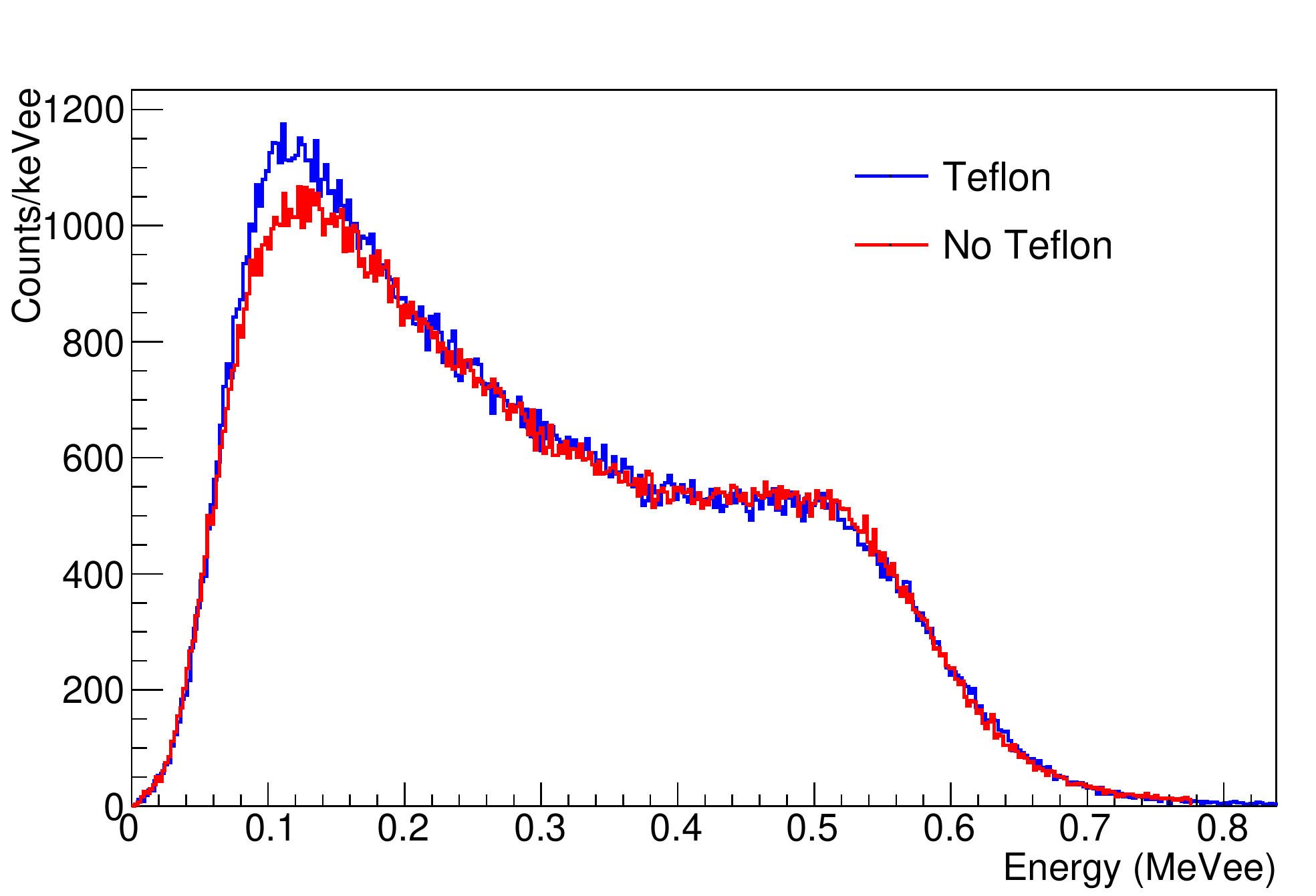}
\caption{Comparison of ${}^{137}$Cs spectra with and without Teflon wrapping of the rods.}
\label{fig_Cs_spectra_teflon_noteflon}
\end{figure}

\section{Performance metrics}
\label{performance}

\subsection{SiPM signal characteristics and amplification}
\label{AmpFastSlow}

Despite the lack of meaningful PSD information available from the fast-output boards, satisfactory energy sensitivity was nevertheless obtained. 
Fig.~\ref{fig_Cs_spectra_comparison_1} shows a series of spectra obtained for ${}^{137}$Cs --- with and without amplifiers, and utilizing both slow and fast outputs.
The best performance, which we define here as the sharpest Compton edge, 
was achieved using slow-output readout and amplifiers (eight 16-channel, 10$\times$-gain fast-amplifier CAEN N979 modules), though the improvement over fast readout was almost negligible.

Another test of the calibration is to measure the relative amount of signal observed in SiPM A and SiPM B as a function of $z$ position. After calibration, events that occur at $z=0$ should produce an equivalent signal at SiPM A and B. The ratio $R_{AB}$, which we refer as the  {\it AB ratio}, is defined as
\begin{equation} \label{eq_AB_ratio}
    R_{AB} \equiv \frac{Q_{A}}{Q_{A} + Q_{B}},
\end{equation}
where $Q_A$ and $Q_B$ represent the charge collected on SiPM A and B respectively. In the following, $R_{AB}$ will be used to determine the $z$ position of each event. In order to maximize position sensitivity, the AB ratio was calculated only for the rod with the largest energy deposit. Fig.~\ref{fig_AB_ratio_vs_energy} represents the AB ratio distributions for a $^{137}$Cs source collimated at $z=0$, for two different energy ranges.

\subsection{Teflon wrapping of scintillator}
\label{sub_teflon_wrap}

We also experimented with wrapping the scintillator bars with Teflon in an attempt to maximize light output. Although the light yield increased slightly when wrapped, we found that the light transport as a function of distance along the scintillator rods suffered. Fig.~\ref{fig_AB_ratio_vs_position} illustrates this effect, for both amplified and non amplified output data, with and without Teflon. The two flattest curves, indicating minimum signal loss with rod length, were from non-Teflon-wrapped data runs. 
We do not know the cause of this effect, however, one possibility might be that the Teflon slightly impacts the effective total internal reflection (TIR) along each scintillator rod.
Since the aspect ratio of the scintillator rods is large, the TIR photons reflect multiple times before reaching a SiPM. If a photon leaks from a rod due to imperfect TIR, it undergoes diffuse reflection off the Teflon and cannot resume its former TIR trajectory. These photons eventually absorb after many reflections. The use of Teflon seems to make no difference to the energy resolution, as shown in Fig.~\ref{fig_Cs_spectra_teflon_noteflon}.

\begin{figure}[ht]
\centering\includegraphics[width=1.0\linewidth]{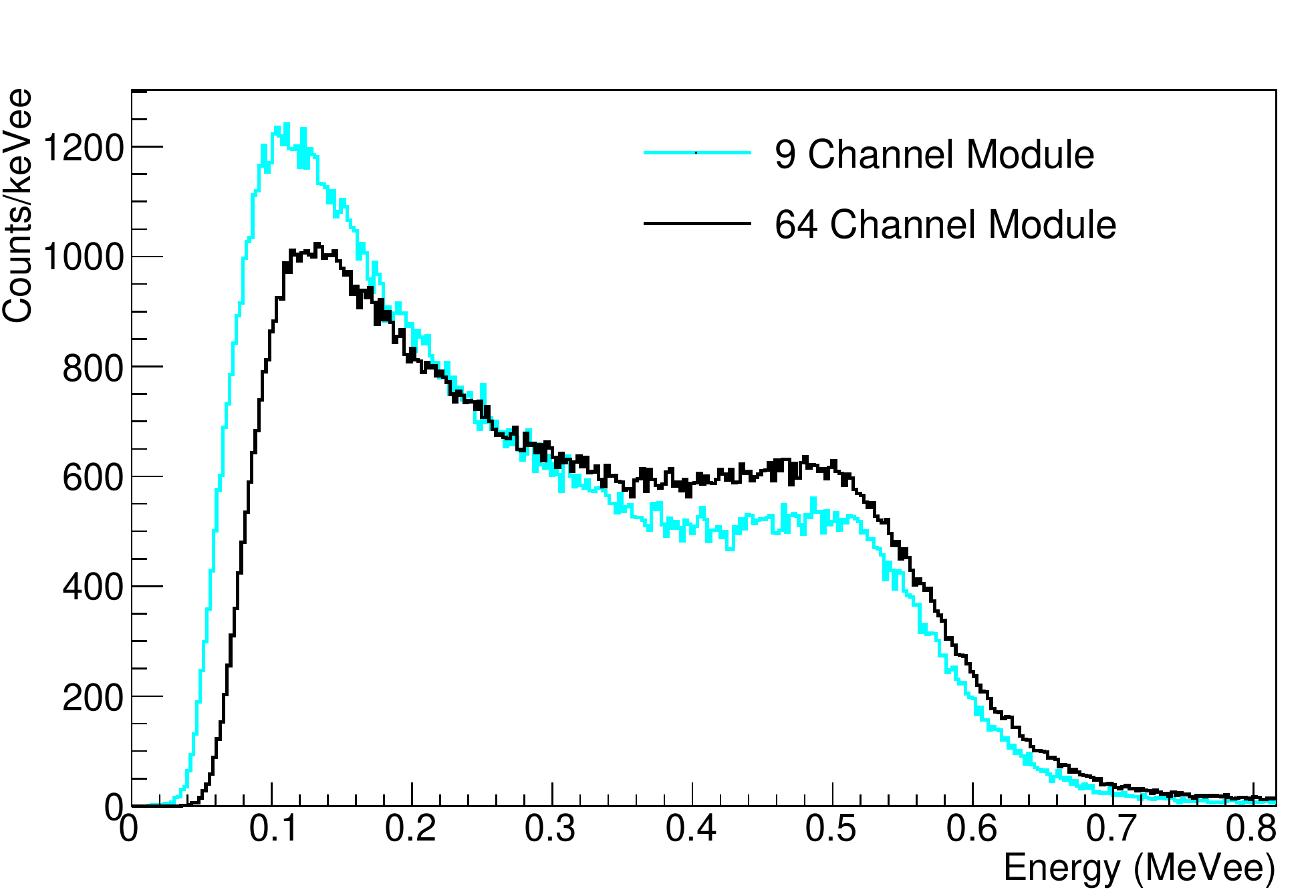}
\caption{Comparison of ${}^{137}$Cs spectra with 9 vs 64 rods.}
\label{fig_Cs_spectra_9vs64rods}
\end{figure}

The tests to investigate effects of the readout boards, amplification and Teflon wrapping were all performed on a 9-rod version of the prototype detector for the sake of convenience. Once the optimal configuration was found, 
the prototype was filled with the remaining rods of the 64-rod assembly. The optimal configuration was to bias the SiPMs at a slightly lower voltage while using amplifiers, and to use non-Teflon-wrapped rods (to maximize light transport along each rod). We will show below the effect of using the slow-output interconnect cards for PSD. We note that the full 64-rod assembly had a more prominent $^{137}$Cs Compton edge than the 9-rod configuration, as shown in Fig.~\ref{fig_Cs_spectra_9vs64rods}. This is because each rod is situated $\lesssim$1~mm from its neighboring rods. Compton-scattered electrons that exit one rod immediately enter a neighboring rod, ensuring that little of the deposited energy is lost, contributing to the Compton edge.

Another important issue to consider is that in some cases we observed an ingress of the optical grease between the scintillator and Teflon tape, near the ends of each rod, further deteriorating the TIR mechanism and subsequently compromising the detector's overall performance. 

\subsection{Position resolution using relative charge (AB ratio)}
\label{subsection_position_resolution}

\begin{figure}[ht]
\centering\includegraphics[width=1.0\linewidth]{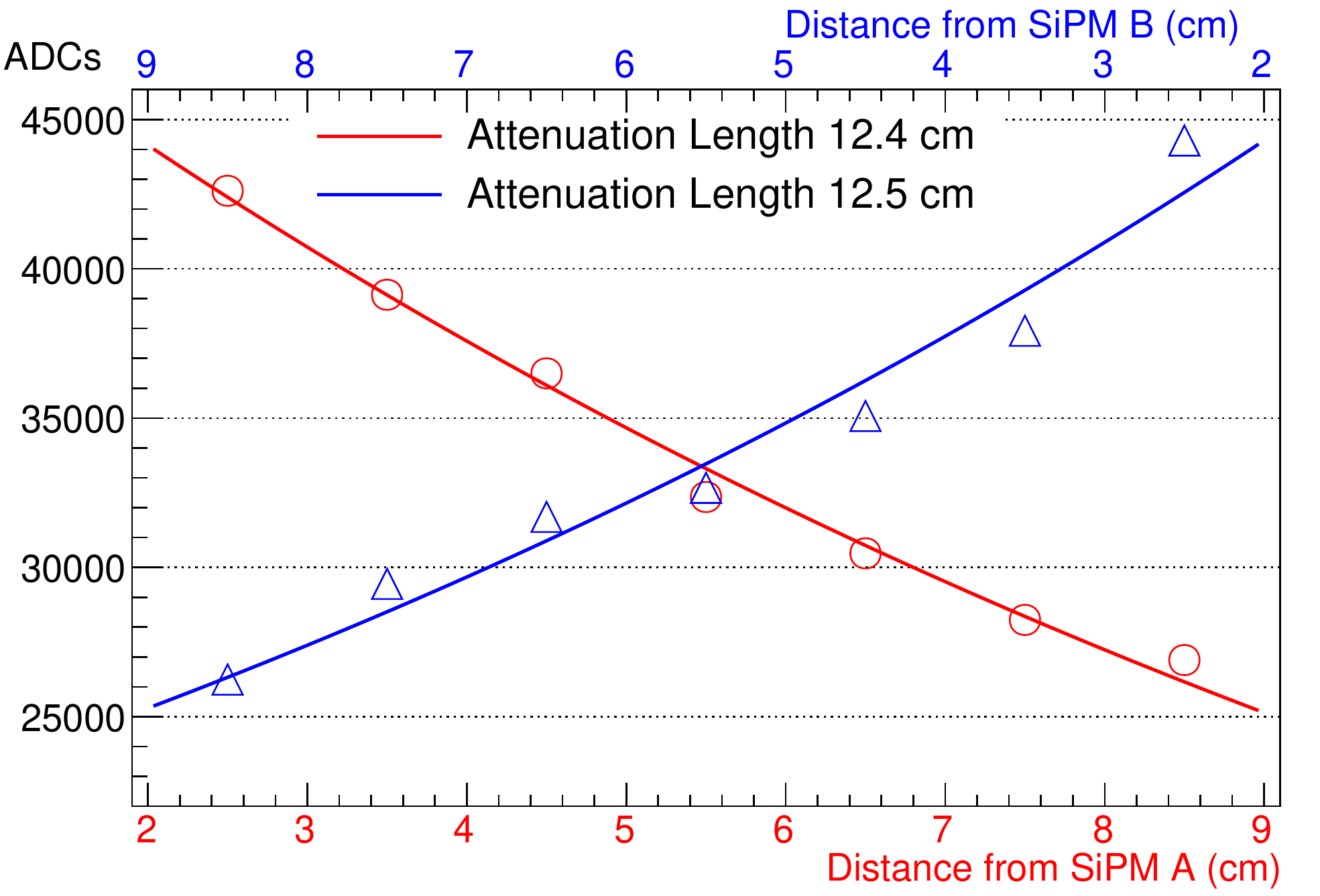}
\caption{Total charge collected on SiPM A (B) as a function of distance from the SiPM A (B).}
\label{fig_att_length_fit}
\end{figure}

In Fig.~\ref{fig_att_length_fit} the effective attenuation length, which includes the effects of the intrinsic attenuation length of the scintillator and reflectivity loss at the surface of each rod segment, was calculated from the exponential fit to the light output as a function of $z$ position from each of the SiPMs channels. 

\begin{figure}[ht]
\centering\includegraphics[width=1.0\linewidth]{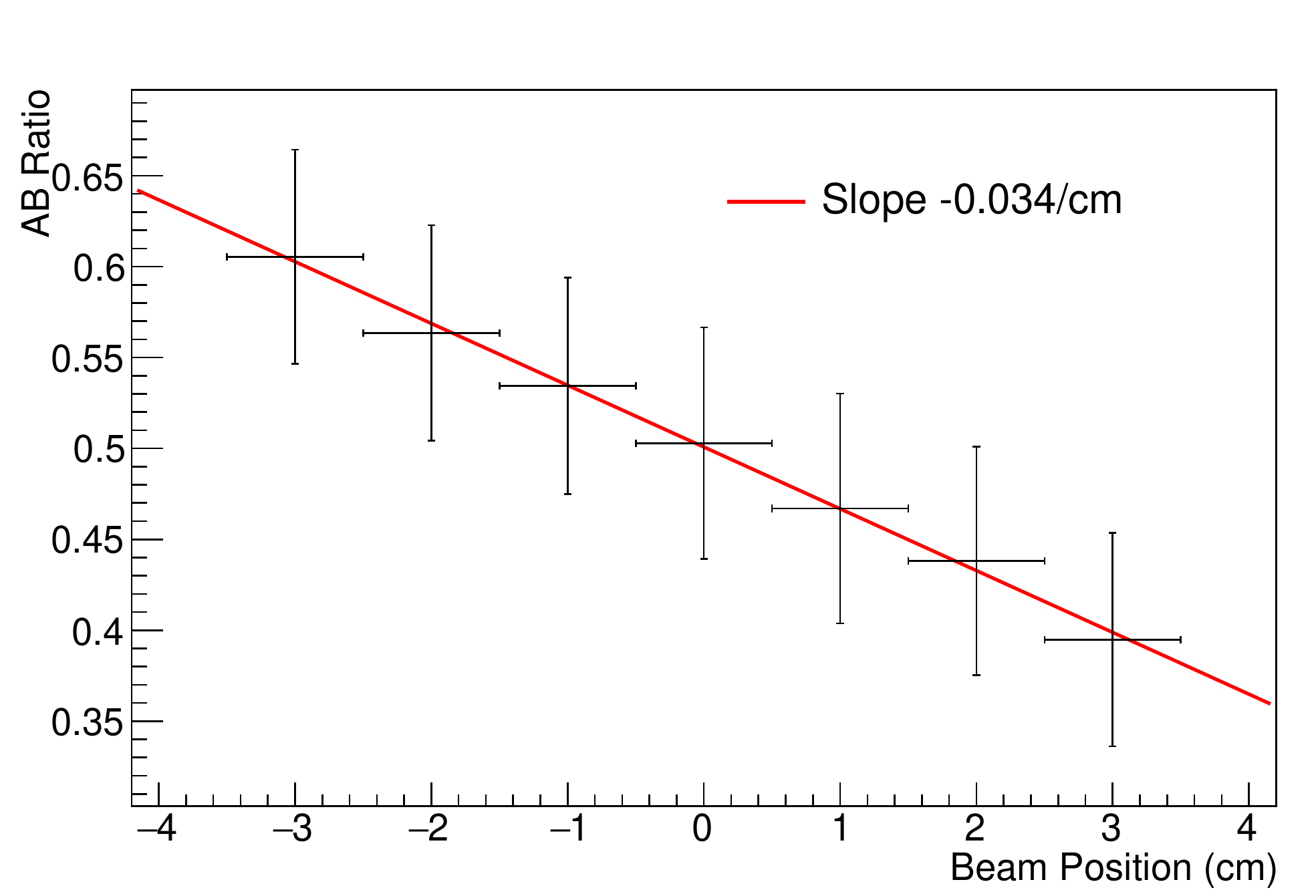}
\caption{The fraction of total charge in SiPM A as a function of $z$ position (AB ratio) using collimated beams of 662-keV gamma rays from a $^{137}$Cs source located at various $z$ positions. Note: The error bars indicate the width of the Gaussian fit to the AB ratio curve (see Fig.~\ref{fig_AB_ratio_vs_energy}), indicating the position resolution on an event-by-event basis, not the uncertainty of the mean.}
\label{fig_ABVersusPosition}
\end{figure}

\begin{figure}[ht]
\centering\includegraphics[width=1.0\linewidth]{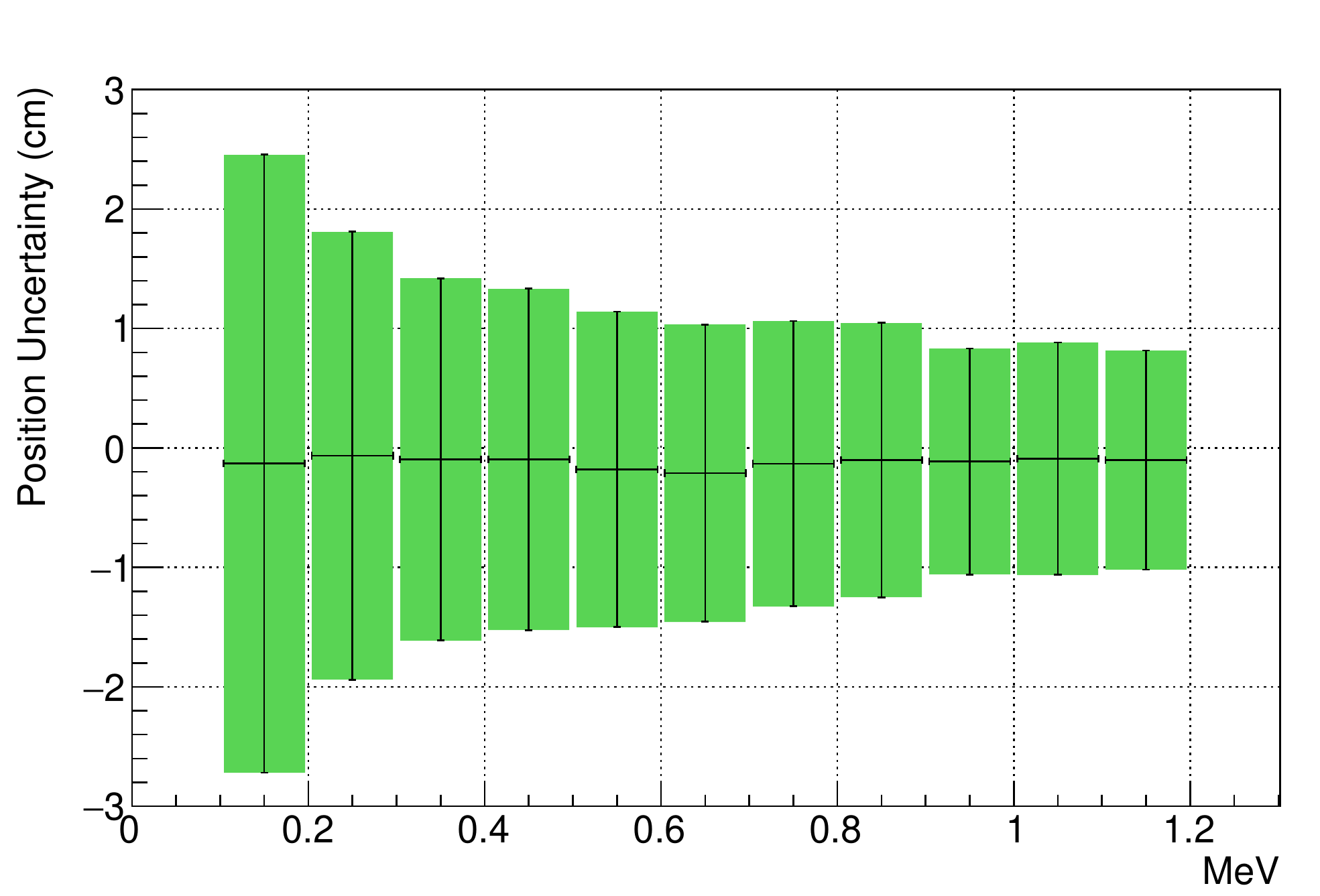}%Cs137_ABResolutionVersusEnergy.pdf}
\caption{Position resolution as a function of energy obtained using a collimated $^{22}$Na gamma-ray source aligned with the center of the detector ($z=0$).}
\label{fig_position_resolution_vs_energy}
\end{figure}

Fig.~\ref{fig_ABVersusPosition} shows a series of measurements of the AB ratio as a function of the $z$ position using collimated gamma rays from a $^{137}$Cs source. A linear fit is shown with a slope of $-0.034$/cm. We use this fit to determine the $z$ position of each event in the detector. The position uncertainty as a function of energy is also shown in Fig.~\ref{fig_position_resolution_vs_energy}. We find the position uncertainty to be $\sim$1~cm at 1~MeV$_{\mathrm{ee}}$. Some continued slight reduction of uncertainty can be anticipated for higher energies. 

\section{Particle identification}
\label{ParticlID}
\subsection{Pulse-shape discrimination}
\label{PSD}

Sensitivity to differences between neutron and gamma-ray interactions via differences in signal pulse shape is an important metric for antineutrino detection and background reduction. 
%In organic scintillators, neutron interactions such as proton recoils and neutron capture on $^{6}$Li produce higher concentrations of molecular triplet states, which yield a longer light pulse than electrons~\cite{ZAITSEVA201288,ZAITSEVA2013747}. %minimally-ionizing 
In this work, relative differences in pulse shape were quantified using the ratio of the tail charge to total charge, as shown in Fig.~\ref{fig:sample_waveform_slowoutput}. The timing cuts were determined via an optimization using a figure-of-merit (FOM) calculated over a broad range of potential $Q_{tail}$ and $Q_{total}$ time ranges. The FOM was defined as
\begin{equation}\label{eq_FOM}
    \mathrm{FOM} \equiv \frac{\mu_{e} - \mu_{n}}{FWHM_{e}+FWHM_{n}},
\end{equation}
where $\mu_{e}$($\mu_{n}$) and $FWHM_{e}$($FWHM_{n}$) are the mean and full width half maximum values of $Q_{tail} / Q_{total}$, for electrons and neutrons, respectively. The resulting optimal time ranges for tail-charge and total-charge integration times are given in Eq.~(\ref{IntegrationTiming}) and Eq.~(\ref{IntegrationTiming2}). 

\begin{figure}[ht]
%\centering\includegraphics[width=1\linewidth]{figs/pspPlot-1.pdf}
\centering\includegraphics[width=1.0\linewidth]{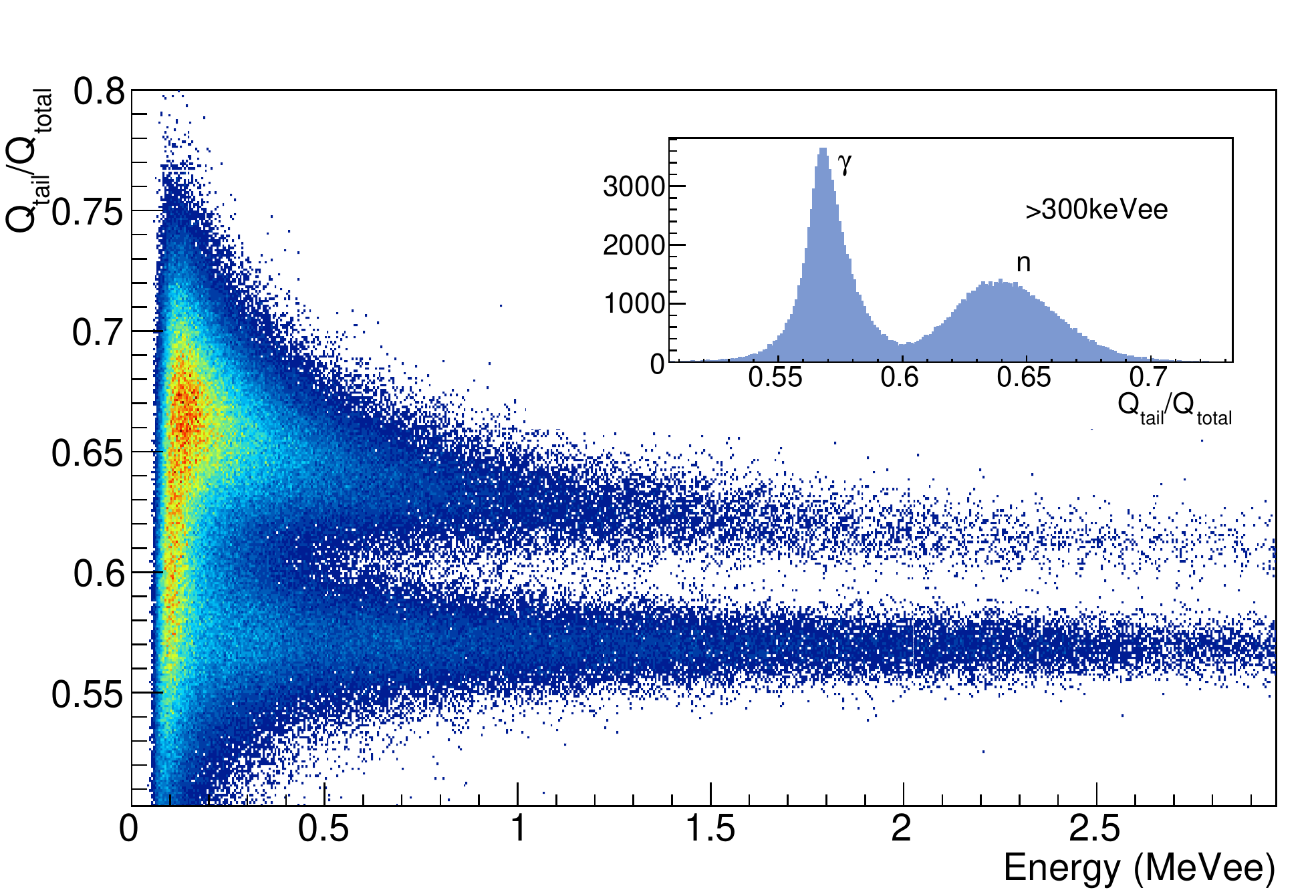}%pspPlot_new.pdf 
\caption{Neutron/gamma pulse-shape discrimination from a ${}^{252}$Cf source, based on tail/total charge analysis using the full 64-rod module. Main plot: a distribution of events as a function of $Q_{\mathrm{tail}}/Q_{\mathrm{total}}$ and total energy. Inset plot: a distribution of $Q_{\mathrm{tail}}/Q_{\mathrm{total}}$ with total energy above 300 keV$_{\mathrm{ee}}$. }
\label{fig_n_gamma_pulse_shape_1}
\end{figure}

\begin{figure}[ht]
%\centering\includegraphics[width=1\linewidth]{figs/pspPlot-1.pdf}
\centering\includegraphics[width=1.0\linewidth]{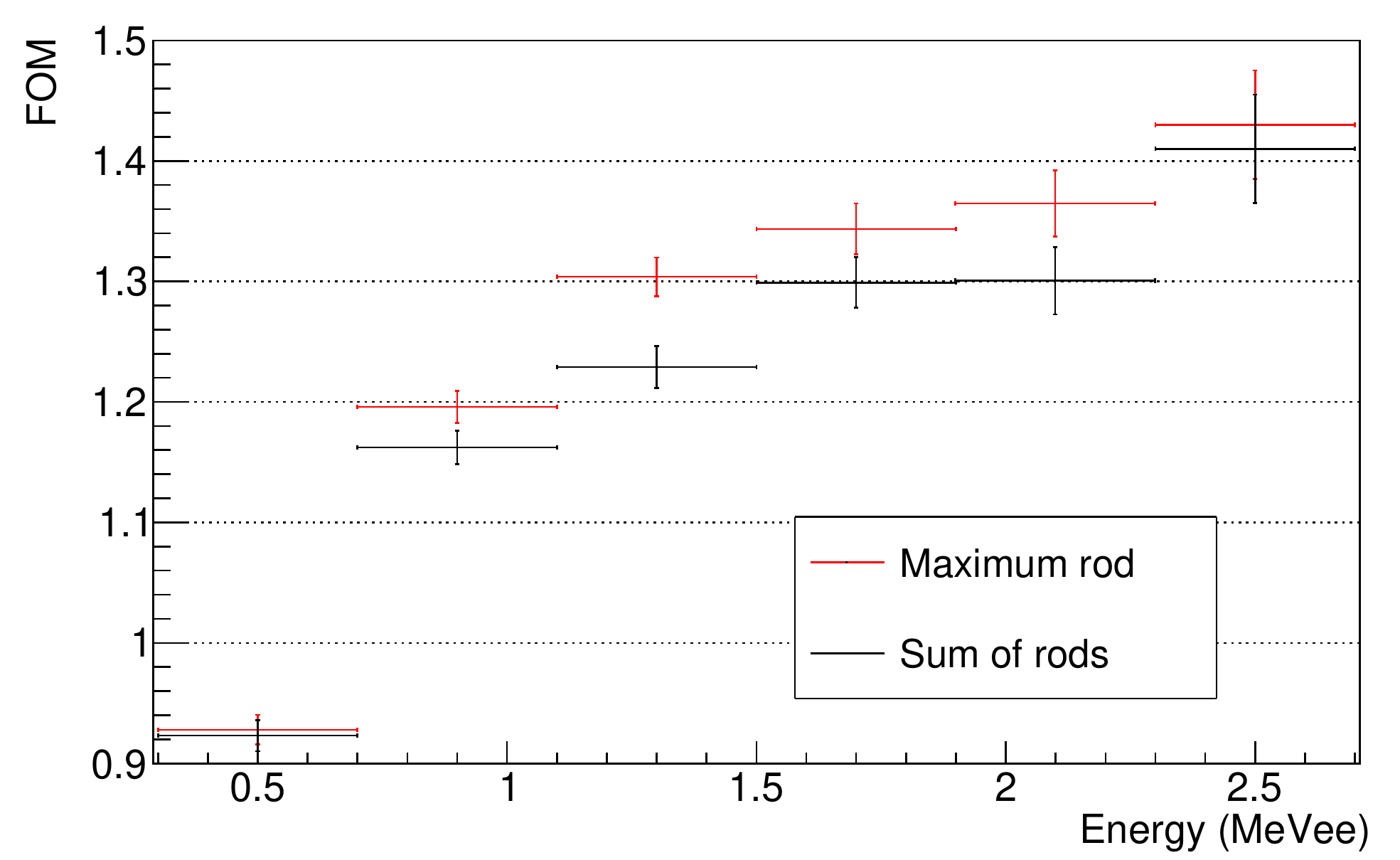} 
\caption{The figure of merit (FOM) as defined in Eq.~\ref{eq_FOM} for discriminating between gamma-rays and neutron scatters in the prototype detector for a variety of energies. Black --- the PSD distribution calculated from all the hit rods (as in Fig.~\ref{fig_n_gamma_pulse_shape_1}); red --- the PSD distribution obtained from the single rod containing the maximum signal in each event.}
\label{fig_FOM}
\end{figure}

Fig.~\ref{fig_n_gamma_pulse_shape_1} shows $Q_{tail}$ / $Q_{total}$ after optimization as a function of electron-equivalent energy for a % and~\ref{fig_n_gamma_pulse_shape_2}.
6-$\mu$Ci ${}^{252}$Cf source shielded behind 15 cm of lead to reduce the gamma-ray component of the flux. 
On average, per fission, ${}^{252}$Cf  emits about 7.8 gammas and 4 neutrons~\cite{Tanabashi:2018oca}. Mean gamma energy is 0.88~MeV; neutron --- 2.14~MeV.
The source was located at $z=0$, 20~cm from the detector. Also shown in the inset figure is a 1-D plot of $Q_{tail}$ / $Q_{total}$ integrated over all energies greater than 300~keV$_{\mathrm{ee}}$.
The optimized FOM as a function of energy is shown in Fig.~\ref{fig_FOM}. 
The visible energy for neutron capture is in  the range 300$-$500~keV$_{\mathrm{ee}}$ for $^6$Li-doped plastic scintillators~\cite{ZAITSEVA2013747,MABE201680}; thus, it is important to achieve good neutron/electron PSD in that energy range.
Moreover, we also note that the PSD is better if only one rod with a maximum light output for a particular event used in the analysis, shown in red points in Fig.~\ref{fig_FOM}.

We note here that these results may represent the first published report of pulse-shape sensitivity from fully-instrumented 64-channel SiPM arrays coupled to plastic scintillator. Until now, most 64-channel readout boards have been of the fast output (differential) variety~\cite{Sweany:2019sxg}. PSD in plastic scintillator or stilbene, using slow (non-differential) readout boards, has been previously reported in single-channel or summed SiPM readouts~\cite{RUCH20151,LIAO2015150,TAGGART2018148}. 
For SANDD, the full 64-channel SiPM pulse-shape sensitivity reported here appears to be adequate for antineutrino detection. However, improvements may be achievable in the future with some further optimization of the readout.

\subsection{Rod multiplicity}
\label{rodMultiplicity}

\begin{figure}[ht!]
	\begin{center}
    \includegraphics[width=0.48\textwidth]{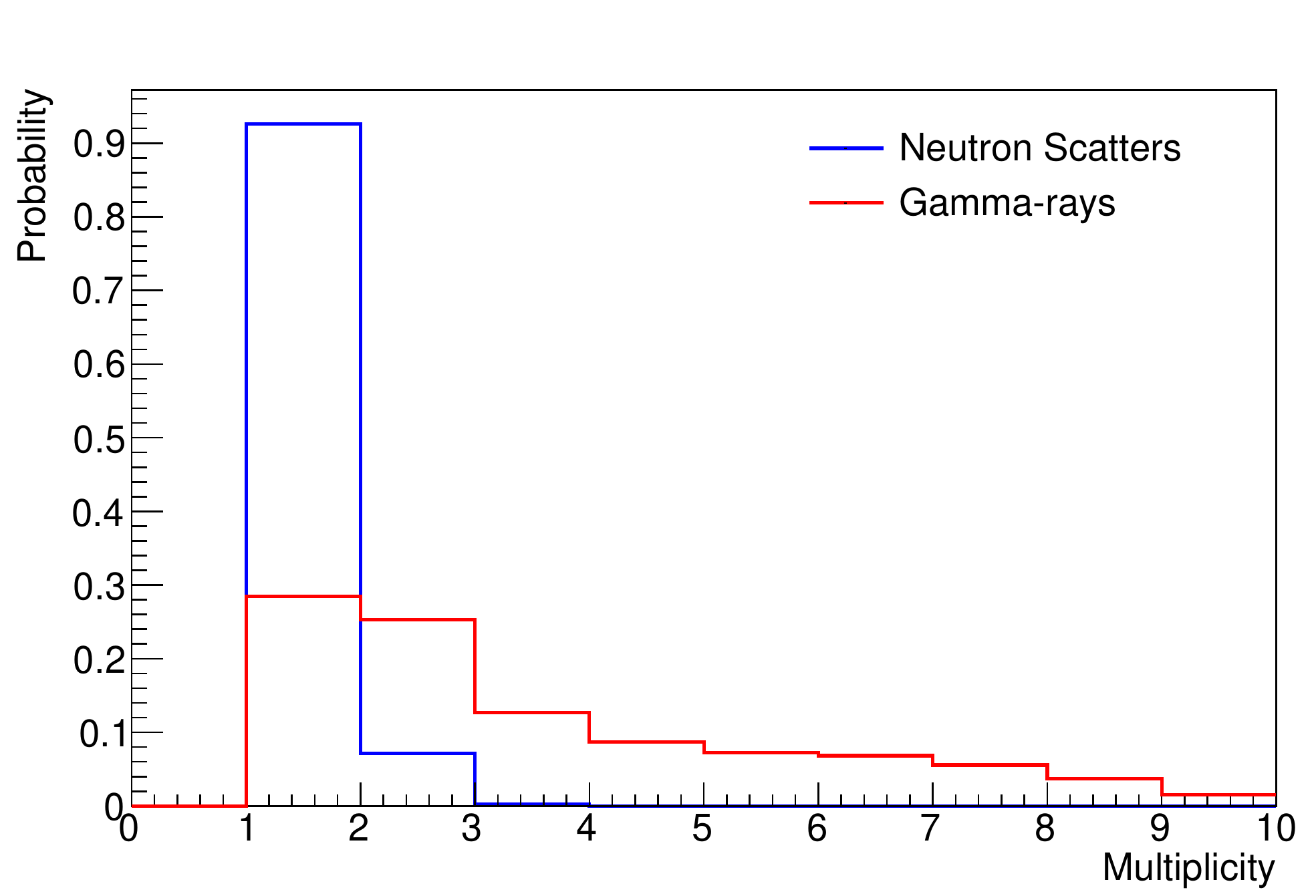}
    {\footnotesize (a)} \\
    \includegraphics[width=0.48\textwidth]{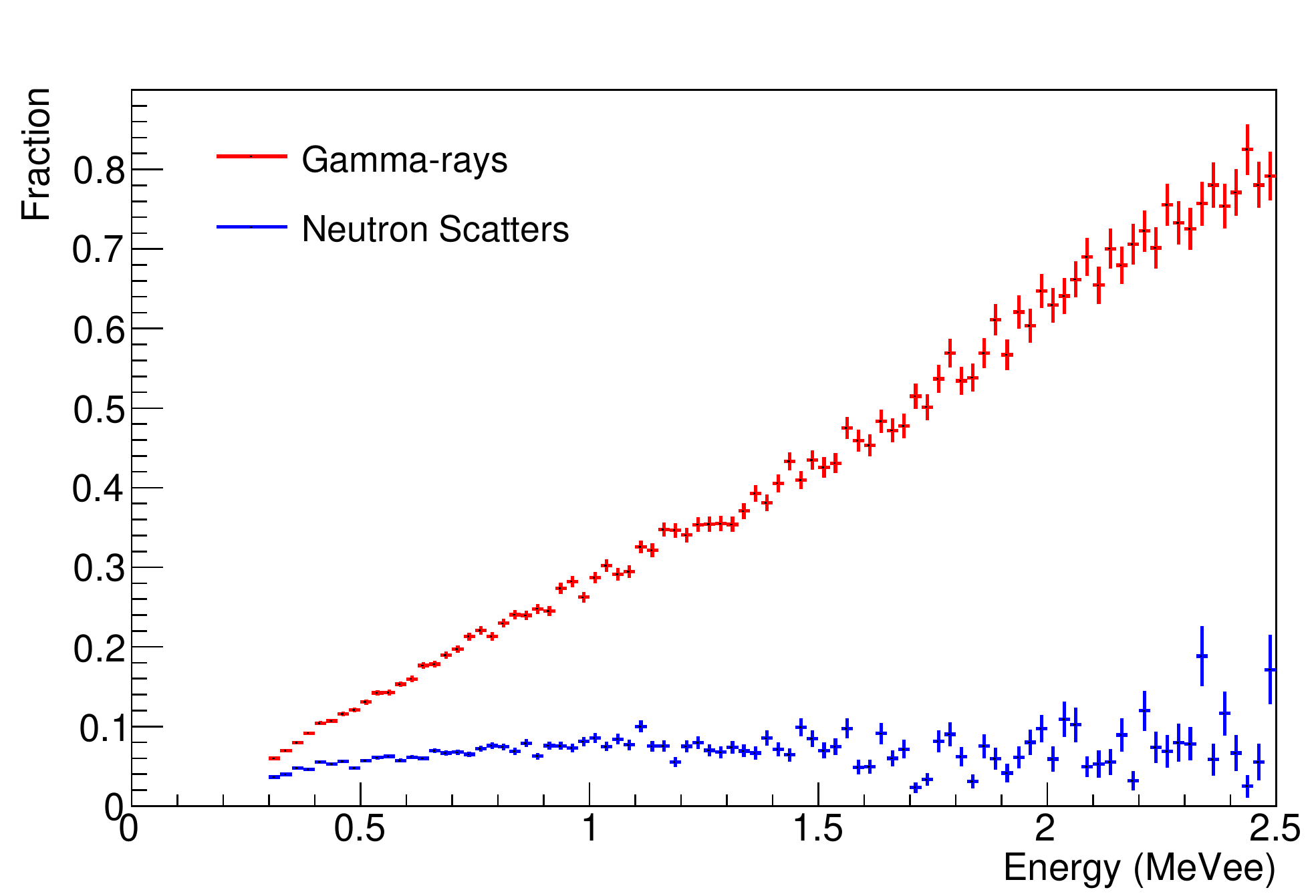}
    {\footnotesize (b)}
	\end{center}
    \caption{(a) Probability distribution of rod multiplicity for gammas and neutrons (for energies $>$1~MeV$_{\mathrm{ee}}$). (b) The fraction of events with multiplicity$\geq$2 as a function of energy (for energies $>$300~keV$_{\mathrm{ee}}$). The neutrons and gamma rays were selected on the basis of their pulse shape from $^{252}$Cf data:
$0.55 < Q_{\mathrm{tail}}/Q_{\mathrm{total}} < 0.59$ -- gamma rays, 
$0.61 < Q_{\mathrm{tail}}/Q_{\mathrm{total}} < 0.68$ -- neutron scatters.}\label{fig_multiplicity_new}
\end{figure}

The PSD parameter in Fig.~\ref{fig_n_gamma_pulse_shape_1} was used to select separate populations of gamma-ray- and neutron-scattering-induced events for this study. At energies greater than $\sim$1~MeV$_{\mathrm{ee}}$, the population of gamma-ray events tends to deposit their energy across more than one scintillator rod, as illustrated in  Fig.~\ref{fig_multiplicity_new}(a). 
This is because electrons deposit their energy over a longer distance in the scintillator than protons, thus traversing multiple rods.
The population of neutrons tends to primarily deposit their energy within a single rod. % ($>$90\%). 
This multiplicity effect will also be used to further discriminate between neutron and gamma events in the SANDD detector.

Fig.~\ref{fig_multiplicity_new}(b) shows the fraction of neutron- and gamma-like events with rod  multiplicity$>$1.
Most neutron-like events have rod multiplicity$=$1 (92\%), which is fairly constant with energy. In contrast, the gamma rays produce higher multiplicity events more often, and at a rate that appears to depend approximately linearly with energy over the range of energies examined here\footnote{It is likely that cosmogenic muons are some fraction of those high-multiplicity gamma-like events at high energies.}.

\section{Conclusion}

A fully-instrumented 64-rod prototype detector, instrumented with two 64-channel SiPM arrays and a 128-channel individual-waveform readout system, has been built. A special PSD plastic scintillator was synthesized as a part of this effort. 
The detector's full 128-channel readout produces neutron/gamma PSD sufficient for use as an antineutrino detector. 
The FOM at particle energies $\gtrsim$300~keV$_{\mathrm{ee}}$ is equal to 1.1 for this prototype detector.

We performed a variety of tests with different configurations and using different radioactive sources. 
We tested individual large aspect ratio Teflon wrapping, amplification combined with lower SiPM bias voltage, differential and non-differential SiPM readout. Our tests showed that Teflon-wrapping is not necessarily the ideal way to maximize light transport along each of the high-aspect-ratio scintillator rods. 

\begin{figure}[ht]
\centering\includegraphics[width=0.75\linewidth]{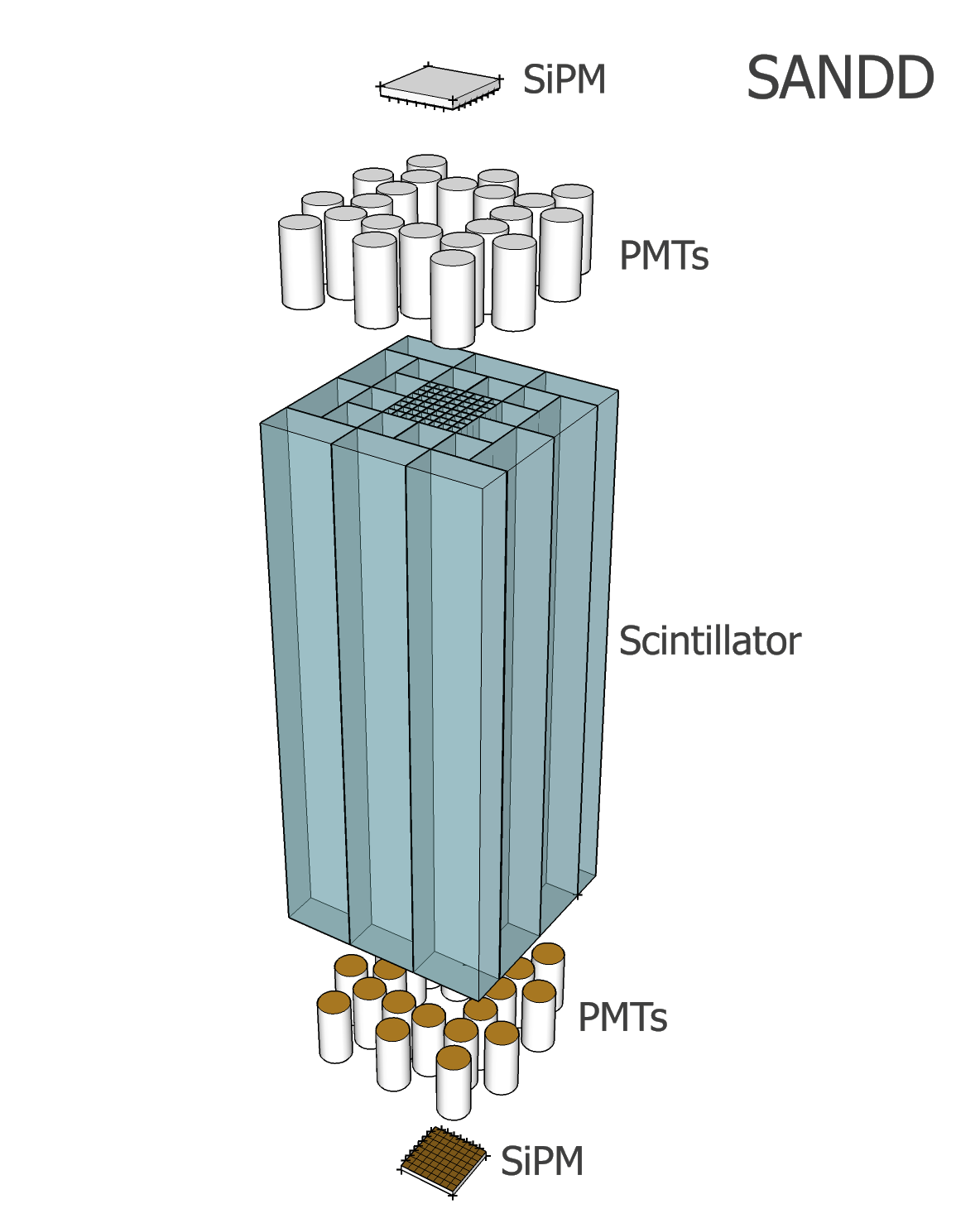}
\caption{Exploded view of the final configuration to be constructed and deployed near a nuclear reactor.}
\label{SANDD_final}
\end{figure}

In the near future, we plan to upgrade the plastic scintillator to a ${}^{6}$Li-loaded formulation at length scales approaching 40 to 50~cm. 
After that, we plan to surround this module with larger scintillator bars of cross-sections 2.5~cm~$\times$~2.5~cm and 2.5~cm~$\times$~5~cm, in readiness for a deployment at a nuclear reactor. The full detector (SANDD) will contain approximately 10~liters of ${}^{6}$Li-doped PSD plastic scintillator, and will be a hybrid, instrumented with both SiPM arrays and PMTs, as shown in Fig.~\ref{SANDD_final}.

To the best of our knowledge, this prototype is the first experiment to report PSD in a fully instrumented SiPM array coupled to plastic scintillator. This work may open new possibilities in the field of compact neutron-imaging applications and reactor-antineutrino directional detectors.

\section{Acknowledgements}

The authors wish to acknowledge Natalia Zaitseva of LLNL for helpful discussions regarding scintillator development.
We also thank \href{http://www.ultralytics.com/}{Ultralytics} and \href{https://www.struck.de/}{Struck} engineers for their support and useful discussions.
The research of F. Sutanto was performed under the appointment to the Lawrence Livermore Graduate Scholar Program Fellowship. The work of I.J. and F.S. was partially supported by the Consortium for Verification Technology under U.S. Department of Energy National Nuclear Security Administration award number DE-NA0002534.
This work was supported by the U.S. Department of Energy National Nuclear Security Administration and Lawrence Livermore National Laboratory [Contract No. DE-AC52-07NA27344, LDRD tracking number 17-ERD-016, release number LLNL-JRNL-769666].

\bibliographystyle{model1-num-names}

%\section{References}
\bibliography{refs_href.bib}

\end{document}